\documentclass[aps,prx,twocolumn,longbibliography]{revtex4-2}

\usepackage{amsmath}
\usepackage{amssymb}
\usepackage{graphicx}
\usepackage{bm,hyperref}
\usepackage{braket}
\usepackage{xcolor}
\usepackage{array}
\usepackage{bbold}
\usepackage{multirow}
\newcolumntype{L}[1]{>{\raggedright\let\newline\\\arraybackslash\hspace{0pt}}b{#1}}
\newcolumntype{C}[1]{>{\centering\let\newline\\\arraybackslash\hspace{0pt}}b{#1}}
\newcolumntype{R}[1]{>{\raggedleft\let\newline\\\arraybackslash\hspace{0pt}}b{#1}}
\renewcommand{\Re}{\textrm{Re}}

\def\vec{\mathbf}

\begin{document}
\title{Stoner ferromagnetism in a momentum-confined interacting 2D electron gas}

\date{\today}
\author{Ohad Antebi}
\affiliation{Department of Condensed Matter Physics, Weizmann Institute of Science, Rehovot 76100, Israel}
\author{Ady Stern}
\affiliation{Department of Condensed Matter Physics, Weizmann Institute of Science, Rehovot 76100, Israel}
\author{Erez Berg}
\affiliation{Department of Condensed Matter Physics, Weizmann Institute of Science, Rehovot 76100, Israel}

\begin{abstract}
In this work we investigate the ground state of a momentum-confined interacting 2D electron gas, a momentum-space analog of an infinite quantum well. The study is performed by combining analytical results with a numerical exact diagonalization procedure. We find a ferromagnetic ground state near a particular electron density and for a range of effective electron (or hole) masses. We argue that this observation may be relevant to the generalized Stoner ferromagnetism recently observed in multilayer graphene systems. The collective magnon excitations exhibit a linear dispersion, which originates from a diverging spin stiffness. 

\end{abstract}

\maketitle

\emph{Introduction.--}
The ground state of a 2D electron gas is determined by a competition between kinetic and interaction energies. In the absence of external fields or ionic potentials, numerical calculations\cite{Ceperley1980-zz,Tanatar1989-ug,Rapisarda1996-om,Attaccalite2002-pr,Drummond2009-be} suggest the ground state is either a paramagnetic Fermi liquid or a Wigner crystal\cite{Wigner1934-dm}, depending on the density, while a Stoner ferromagnetic\cite{Stoner1938-tg} Fermi liquid state is a close competitor. By applying an out-of-plane magnetic field, the energy spectrum of the electron gas forms Landau levels. In the absence of Zeeman coupling the ground state is known to be ferromagnetic for densities near one electron per flux quantum\cite{Nomura2006-tk,Girvin1999-rt}.

\begin{figure}[t]
    \centering
    \begin{tabular}{ C{0.11\textwidth} | C{0.12\textwidth}  C{0.12\textwidth}  C{0.12\textwidth} }
     &
    Low-density semiconductor &
    Rhombohedral multilayer graphene &
    Landau levels \\
    \hline
    {Single-particle dispersion}
    & \begin{minipage}{0.12\textwidth}\includegraphics[width=\textwidth]{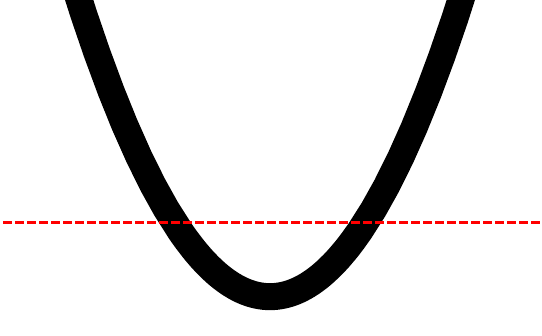}
    \end{minipage}
    & \begin{minipage}{0.12\textwidth}\includegraphics[width=\textwidth]{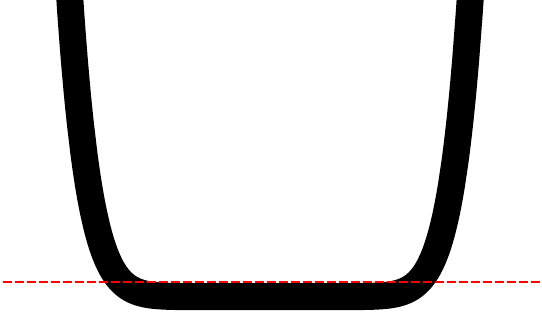}
    \end{minipage} 
    & \begin{minipage}{0.12\textwidth}\includegraphics[width=\textwidth]{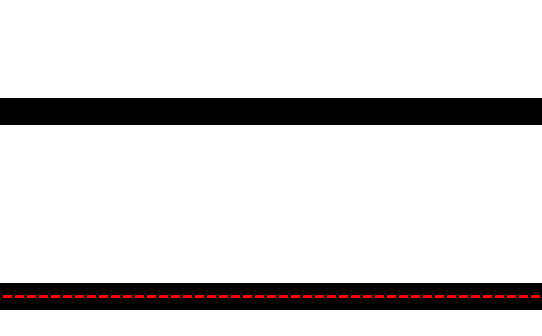}
    \end{minipage} \\
    \hline
    Stoner Magnetism & No* & Yes** & Yes
    \end{tabular}
    \caption{Comparison of dispersion and Stoner ferromagnetism between different 2D systems. The dashed horizontal line represents the chemical potential. *Based on state of the art quantum Monte Carlo studies~\cite{Drummond2009-be}. **Experimentally known for bilayer\cite{zhou2022isospin,zhang2023enhanced,de2022cascade,seiler2022quantum} and trilayer\cite{Zhou2021-tp} graphene, both with a strong perpendicular displacement field, and analytically shown in this work for infinitely many layers.}
    \label{fig:schematic_dispersion}
\end{figure}

Recent experiments in Bernal bilayer and rhombohedrally stacked multilayer graphene show spin and valley ferromagnetic phases for certain regimes of electron density and out-of-plane displacement field\cite{Zhou2021-tp,zhou2022isospin,zhang2023enhanced,de2022cascade,seiler2022quantum,shi2020electronic}. Furthermore, superconductivity has been experimentally demonstrated for the bilayer in Refs.~\cite{zhou2022isospin,zhang2023enhanced,Holleis2023} and the trilayer in Ref.~\cite{zhou2021superconductivity}, and theoretically discussed for various graphene multilayers in Refs.~\cite{awoga2023superconductivity,chatterjee2022inter,Chou2021acoustic,You2022Kohn,Ghazaryan2021uncon,Szabo2022Metals,Cea2022super,Lu2022corr,Ghazaryan2023multilayer,Kopnin2013hight,Munoz2013tb,Ojajarvi2018comp}.  The band dispersion of  the rhombohedrally stacked multilayer graphene systems is approximately flat up to some momentum scale, at which point the kinetic energy increases rapidly with momentum. This corresponds to a high density of states in a bounded region of momentum space, and is an example of a partially flat band. Here, we argue that such dispersion is favorable for Stoner ferromagnetism when the flat region is nearly fully occupied. We summarize these results in Fig.~\ref{fig:schematic_dispersion}. Examples of various correlated phenomena that had been previously studied on models with partially flat bands can be found in Refs.\cite{Mielke1999-zq,Tang2014-ph,Huang2019-zj,Sayyad2020-uu,Aoki2020-og,Sayyad_2023}. 

Motivated by multilayer rhombohedrally stacked graphene, we present a toy model for an interacting 2D electron gas in which the kinetic energy of the electrons diverges beyond a limited region in momentum space. We will refer to this as a momentum-confined gas. We find the ground state of this gas to be spin-polarized. We find a linearly dispersing magnon branch of excitations, in contrast with the commonly predicted quadratic dispersion\cite{Watanabe2020-sz}. We trace this apparent anomaly to a divergence of the spin stiffness associated with the infinite slope of the kinetic energy dispersion. We show that the spin polarization is robust to an addition of weak dispersion to the kinetic energy within the allowed momentum region.

\emph{Hamiltonian.--}
Our goal is to construct a simple model that captures non-trivial interaction effects 
of electrons confined in momentum space,
with a minimal set of parameters. We consider a 2D electron gas with an artificially constructed kinetic dispersion, and turn on the conventional Coulomb repulsion. We imagine the kinetic dispersion to be such that the electrons are only allowed to occupy momenta in a disk of radius $k_0$ in momentum space, whose area is small compared to the Brillouin zone. Idealizing the dispersion, we set the kinetic energy to zero within the disk and to infinity outside it:
\begin{equation}\label{eq:well_dispersion}
    E_{\text{K}}(\vec{k}) =
    \begin{cases}
    0, & |\vec{k}|\leq k_0, \\
    \infty, & |\vec{k}| > k_0.
    \end{cases}
\end{equation}
This dispersion leads to the strict confinement of the electrons to a disk in momentum space, forming a momentum space analog of a 2D circular infinite quantum well in real space. 

The confinement in Eq.~\eqref{eq:well_dispersion} can be thought of as mimicking the low energy band dispersion of $N_{l}$ layers of rhombohedral graphite subject to a displacement field $D$. For this system, there is a band gap proportional to $D$, the dispersion is very flat up to a momentum scale $k_0$, and rises as $E(k)\sim (k/k_0)^{N_{l}}$ for $k>k_0$. We emphasize that we are interested in an isolated band, and therefore consider $D\neq 0$ in this picture. In the limit of $N_{l}\rightarrow \infty$, the scale $k_0$ is set by the ratio of the interlayer tunneling $t_{\perp}$ and the monolayer Dirac velocity $v_{D}$ (see \ref{sec:rhombohedral_graphite} in the Supplemental Material - SM). In this limit, the dispersion resembles the idealized momentum confined model of Eq.~\eqref{eq:well_dispersion}.

\begin{figure}[t]
    \centering
    \includegraphics[width=0.3\textwidth]{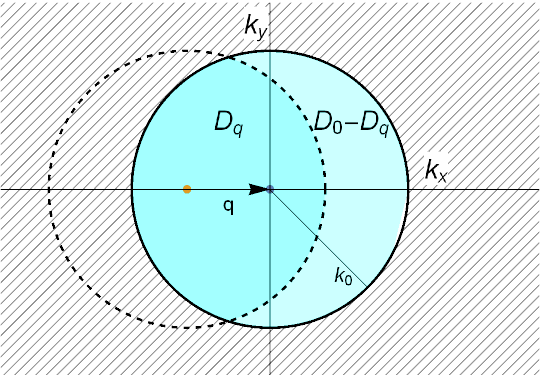}
    \caption{The kinetic dispersion confines electrons to the disk $D_0$ of radius $k_0$ in momentum space. The region $D_{\vec{q}}$ is the intersection of two disks mutually shifted by $\vec{q}$. The strict momentum confinement forbids electrons with momenta in $D_0-D_{\vec{q}}$ from scattering with a momentum transfer $\vec{q}$.}
    \label{fig:dq_region}
\end{figure}

Next, we examine the effects of electron-electron interactions in such a momentum-confined setup. 
We consider our Hamiltonian to be the normal-ordered 2D screened Coulomb interaction strictly confined in momentum space:
\begin{align}\label{eq:single_orbital_hamiltonian}
    \mathcal{H}_{\text{int}} 
    &= \frac{1}{2A}\sum_{\vec{q}}V_{\vec{q}}:\overline{\rho_{\vec{q}}}^{\dagger}\overline{\rho_{\vec{q}}}:, \\
    \overline{\rho_{\vec{q}}} &=  \sum_{\sigma=\uparrow,\downarrow}\sum_{
    \vec{k}\in D_{\vec{q}}}c_{\sigma,\vec{k}+\vec{q}}^{\dagger}c_{\sigma,\vec{k}}. \label{eq:proj_particle_density}
\end{align}
where $c_{\sigma,\vec{k}}$ annihilates an electron with spin $\sigma\in\lbrace\uparrow,\downarrow\rbrace$ and momentum $\vec{k}$, $A$ is the system area, $V_{\vec{q}}=2\pi e^{2}\tanh{(|\vec{q}|d)}/|\vec{q}|$ is the Fourier transform of the Coulomb potential, and $d$ is the distance to the screening gate. 
The sum over momenta in the confined particle density operator, $\overline{\rho_{\vec{q}}}$, is limited to the domain $D_{\vec{q}}=\lbrace \vec{k} \enspace | \enspace(|\vec{k}|\leq k_{0}) \cap (|\vec{k}+\vec{q}|\leq k_{0})\rbrace$ such that the fermion operators are within the disk of radius $k_0$ (see Fig.~\ref{fig:dq_region}). While this Hamiltonian has only two (spin) flavors, some of the conclusions we draw below are applicable also in the presence of multiple valleys. We remark that any physical realization of the kinetic dispersion would be accompanied by form factors in Eq.~\eqref{eq:proj_particle_density} that originate from the Bloch wavefunctions. For simplicity, in this work we set the form factors to one.

The Hamiltonian in Eq.~\eqref{eq:single_orbital_hamiltonian} is the one we will explore for the rest of the paper. It has a continuous rotational symmetry around $\vec{k}=0$, and a global $SU(2)$ symmetry for spin rotations. The most interesting property emerges in the limit of unscreened Coulomb interaction, $\left.k_0 d\rightarrow\infty\right.$. In this limit, the Hamiltonian has only one length scale, given by $l\sim k_0^{-1}$. Consequently, we find a single energy scale $E\sim e^{2}k_0$, and an electron density scale $n_0= k_0^{2}/(4\pi)$, which corresponds to completely filling the disk with a single spin flavor. Defining the filling factor by $\nu=n/n_0$, we will focus our discussion in this paper on $\nu=1$ and on $\nu=1\pm\varepsilon$ for $0<\varepsilon\ll1$.

\emph{The role of normal-ordering.--}
The notion of the single-particle kinetic dispersion in the presence of many-body interactions is not without subtleties. In our model, the flat nature of the dispersion is tied to our choice to consider the normal-ordered form of the interaction operator. Physically, the normal-ordering prevents an electron from interacting with itself. By undoing the normal-ordering of the confined Coulomb interaction, we find that Eq.~\eqref{eq:single_orbital_hamiltonian} can be separated into a positive semi-definite density-density operator and a single-particle operator:
\begin{equation}\label{eq:psd_decomposition}
    \mathcal{H}_{\text{int}} 
    =
    \underbrace{\frac{1}{2A}\sum_{\vec{q}}
    V_{\vec{q}}\overline{\rho_{\vec{q}}}^{\dagger}\overline{\rho_{\vec{q}}}}_{=\mathcal{H}_{\rho-\rho}}
    -\sum_{\sigma=\uparrow,\downarrow}\sum_{|\vec{k}|\leq k_0}
    E_{0}(\vec{k})
    c_{\sigma,\vec{k}}^{\dagger}c_{\sigma,\vec{k}},
\end{equation}
where we introduced the dispersion:
\begin{equation}\label{eq:exact_h_dispersion}
    E_{0}(\vec{k})=\frac{1}{2A}\sum_{|\vec{k}'|\leq k_0}V_{\vec{k}-\vec{k}'}.
\end{equation}
This dispersion is, up to a sign, the exchange contribution to the self-energy of a particle at momentum $\vec{k}$ in the presence of a completely filled disk of radius $k_0$.
In contrast with the standard unconfined interaction, we find that the momentum cutoff introduces a non-trivial dispersion term associated with the normal-ordering. 
\begin{figure}[t]
    \centering
    \includegraphics[width=0.5\textwidth]{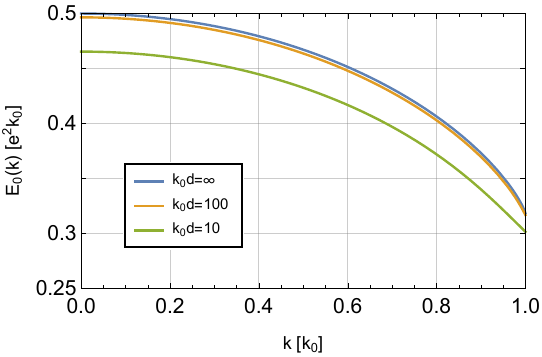}
    \caption{The dispersion $E_{0}(|\vec{k}|)$ for different screening lengths of the Coulomb interaction. In the case of unscreened Coulomb interaction ($k_0d\rightarrow\infty$) the dispersion is singular at $k=k_0$. This singularity is smoothed out for screened Coulomb interaction.}
    \label{fig:e0_vs_k}
\end{figure}
The dispersion in Eq.~\eqref{eq:exact_h_dispersion} plays an important role in several properties of our model and thus warrants an explicit discussion. In the limit of 
 unscreened interaction, $k_0d\rightarrow\infty$, this dispersion is given by:
\begin{equation}\label{eq:unscreened_h_dispersion}
    E_{0}(\vec{k})=e^{2}k_0\frac{1}{\pi}\tilde{E}\left(\frac{\pi}{2},\frac{|\vec{k}|}{k_0}\right)
\end{equation}
where $\tilde{E}\left(\frac{\pi}{2},x\right)$ is the complete elliptic integral of the second kind whose argument $x$ is the elliptic modulus. The derivative of Eq.~\eqref{eq:unscreened_h_dispersion} diverges logarithmically at $|\vec{k}|=k_0$. This divergence is cut off for finite $k_0d$.
Expanding Eq.~\eqref{eq:exact_h_dispersion} in $(k_0 d)^{-1}$, we find that the series does not uniformly converge on the disk, but nevertheless, the correction is small everywhere for $k_0d\gg1$. The effect of screening length on the dispersion is plotted in Fig.~\ref{fig:e0_vs_k}. The details are described in the SM~\ref{sec:normal_ordering}.

\emph{Exact ground state.--}
The ground state of Eq.~\eqref{eq:single_orbital_hamiltonian} is not analytically solvable. We can, however, solve for the exact ground state upon introducing a particular dispersion to the otherwise flat disk, which would eliminate the contribution of Eq.~\eqref{eq:exact_h_dispersion}. With this dispersion, we are left with the density-density Hamiltonian $\mathcal{H}_{\rho-\rho}$ as defined in Eq.~\eqref{eq:psd_decomposition}.  The Hamiltonian $\mathcal{H}_{\rho-\rho}$ is positive semi-definite, which provides a direct route to finding its exact ground state. For $\nu=1$, consider completely filling the spin $\sigma=\downarrow$ flavor, i.e. a maximally spin-polarized state. We denote this state by $\ket{\Psi_{\text{SP}}}$. By applying Eq.~\eqref{eq:proj_particle_density} one easily finds that:
\begin{equation}
    \forall \vec{q}: \overline{\rho_{\vec{q}}}\ket{\Psi_{\text{SP}}} = 0 \implies 
    \mathcal{H}_{\rho-\rho}\ket{\Psi_{\text{SP}}}=0,
\end{equation}
which proves that $\ket{\Psi_{\text{SP}}}$ is a ground state of $\mathcal{H}_{\rho-\rho}$. We emphasize that this ground state is degenerate with $SU(2)$ spin rotation symmetry. For a model with multiple valleys, the above statement holds for any integer filling $\nu$ and any completely spin and valley polarized state, provided that a large momentum separation between different valleys allows for a neglect of inter-valley scattering in Coulomb processes.

The spin polarized state $\ket{\Psi_{\text{SP}}}$ is also an eigenstate of $\mathcal{H}_{\text{int}}$. We \textit{hypothesize} that $\ket{\Psi_{\text{SP}}}$ is a ground state of $\mathcal{H}_{\text{int}}$. We provide numerical evidence for this claim below. Assuming this hypothesis is correct, we proceed to calculate the excitations with respect to this state.

\emph{Single-particle excitations.--}
The single electron and hole excitations energies relative to the fully spin $\sigma=\downarrow$ polarized state, $E_{e}(\vec{k})$ and $E_{h}(\vec{k})$ respectively, are defined by:
\begin{align}\label{eq:sp_excitation_defs}
\begin{split}
    [\mathcal{H}_{\text{int}},c_{\uparrow,\vec{k}}^{\dagger}]\ket{\Psi_{\text{SP}}}
    &= E_{e}(\vec{k})c_{\uparrow,\vec{k}}^{\dagger}\ket{\Psi_{\text{SP}}}, \\
    [\mathcal{H}_{\text{int}},c_{\downarrow,\vec{k}}]\ket{\Psi_{\text{SP}}}
    &= E_{h}(\vec{k})c_{\downarrow,\vec{k}}\ket{\Psi_{\text{SP}}},
\end{split}
\end{align}
and by direct calculation are found to be:
\begin{align}\label{eq:sp_excitation_res}
\begin{split}
    E_{e}(\vec{k}) &= \frac{1}{A}\sum_{|\vec{k}'|\leq k_0}V_{\vec{0}},
    \\
    E_{h}(\vec{k}) &= \frac{1}{A}\sum_{|\vec{k}'|\leq k_0}\left(V_{\vec{k}-\vec{k}'}-V_{\vec{0}}\right).
\end{split}
\end{align}
The identity $E_{0}(\vec{k})=\left(E_{e}(\vec{k})+E_{h}(\vec{k})\right)/2$ holds also if one adds a spin-independent single-particle dispersion to the Hamiltonian $\mathcal{H}_{\text{int}}$.
The electron excitation energy is the charging energy of the system's geometric capacitance, and thus has a flat dispersion. The hole excitation energy has the same contribution (with an opposite sign), along with the exchange interaction of the missing electron. The dispersion of the hole excitation energy is therefore shown in Fig.~\ref{fig:e0_vs_k} up to a factor of two.
We find that the hole dispersion is such that the lowest energy is obtained by removing a hole from the edge of the disk, thereby reducing the Fermi sea radius by an infinitesimal amount. This is identical to the Fermi liquid behavior.

\emph{Collective excitations.--}
The collective particle-hole excitations of momentum $\vec{Q}$ and spin $\hbar$ with respect to $\ket{\Psi_{\text{SP}}}$ are eigenstates of $\mathcal{H}_{\text{int}}$ spanned by wavefunctions of the form:
\begin{equation}\label{eq:col_exc_basis}
    \ket{\Psi_{\text{ph}}^{\vec{k},\vec{Q}}}=c_{\uparrow,\vec{k}+\vec{Q}}^{\dagger}c_{\downarrow,\vec{k}}\ket{\Psi_{\text{SP}}}.
\end{equation}
We remark the state described by $\ket{\Psi_{\text{SP}}}$ does not admit particle-hole excitations that are spinless, in contrast with the Stoner metallic state, and in some similarity to a ferromagnetic band insulator.
The solution is given by diagonalizing the following matrix:
\begin{equation}\label{eq:col_exc_mat_inf}
    [H(\vec{Q})]_{\vec{k},\vec{k}'} = 
        \bra{\Psi_{\text{ph}}^{\vec{k}',\vec{Q}}}\mathcal{H}_{\text{int}}
        \ket{\Psi_{\text{ph}}^{\vec{k},\vec{Q}}}, 
        \quad
        \vec{k},\vec{k}'\in D_{\vec{Q}}, 
\end{equation}
This restriction of both indices to the domain $D_{\vec{Q}}=\lbrace \vec{k} \enspace | \enspace(|\vec{k}|\leq k_{0}) \cap (|\vec{k}+\vec{Q}|\leq k_{0})\rbrace$ is due to our infinite kinetic energy dispersion for states with momentum outside the disk. This sharp cutoff leads to a striking result - the lowest-lying particle-hole excitation is massless. For small $|\vec{Q}|/k_0$ we find:
\begin{equation}\label{eq:linear_ph_disp}
    E_{ph}(\vec{Q}) \approx \frac{4 E_{0}(k_0)}{\pi k_0}|\vec{Q}|.
\end{equation}
This is surprising, as for a magnon we normally expect to find a finite spin stiffness $ \rho_{s}$ such that $E_{ph}(\vec{Q})\approx \rho_{s}|\vec{Q}|^{2}$. The above result holds in the presence of any rotationally symmetric single-particle dispersion, see SM~\ref{sec:collective_ph_excitations}.

To resolve this apparent discrepancy, we replace our strictly confining kinetic dispersion with the following:
\begin{equation}\label{eq:hamiltonian_soft_well}
    \mathcal{H}_{\text{K}}=U_{\text{K}}\sum_{\sigma=\uparrow,\downarrow}\sum_{\vec{k}}\left(\frac{|\vec{k}|}{k_0}\right)^{N_{D}}c_{\sigma,\vec{k}}^{\dagger}c_{\sigma,\vec{k}},
\end{equation}
where $U_{\text{K}}$ is chosen such that $0<U_{\text{K}}
\ll e^{2}k_0$. Under this softened momentum confinement the momentum summations are unconstrained since particles are allowed to be excited beyond the disk. We emphasize that in the limit of $N_{D}\rightarrow\infty$ our strictly confining kinetic dispersion is restored. For $N_{D}\gg 1$ we find, using second order perturbation theory, that the spin stiffness is finite, and the magnon dispersion is quadratic:
\begin{equation}\label{eq:soft_ph_disp}
    E_{ph}(\vec{Q})\approx \frac{1}{2}\frac{U_{\text{K}}}{k_0^{2}}N_{D}|\vec{Q}|^{2}.
\end{equation}
However, in the limit $N_{D}\rightarrow\infty$, the spin stiffness $\left.\rho_{s}\sim N_{D}\right.$ diverges. This leads to a breakdown of perturbation theory as $|\vec{Q}|/k_0$ can no longer be used as a small parameter. The details are described in SM~\ref{sec:collective_ph_excitations}.

\emph{Phase diagram near $\nu=1$.--}
We have seen above that when removing electrons (or adding holes) to the spin-polarized ground state $\ket{\Psi_{\text{SP}}}$, it is energetically favorable to start with the outermost states, thereby reducing the radius of the occupied disk. Consider removing a small fraction of the electrons from the system, i.e. setting the system at filling fraction $\nu=1-\varepsilon$ for $0<\varepsilon\ll1$. We now argue that the system has a spin-polarized Fermi liquid ground state for small enough $\varepsilon > 0$. In the limit of $\varepsilon\rightarrow 0$, the spin polarization is undiminished, and the system can be thought of as having a single species of particles (in this case, holes), with a single particle dispersion $E_{h}(\vec{k})=2E_{0}(\vec{k})$ and subject to the 2D screened Coulomb interaction. Partially filling this system leads to a circular Fermi surface. The finite interaction strength (due to screening) and the constraint to 2D imply a Fermi liquid ground state. Presumably, continuing to decrease the filling will eventually destroy the spin polarization, and we expect a phase transition to a spin-depolarized state.

When the dispersion is exactly flat ($N_{D}\rightarrow\infty$) there is difficulty making a similar argument for filling $\nu=1+\varepsilon$, as the single-electron excitation spectrum is completely flat. However, as previously discussed, the average of single-electron and single-hole excitation spectra is constrained to be $E_{0}(\vec{k})$. Therefore, upon introduction of some weak dispersion to the disk, for example - by setting $N_{D}$ to a finite value, we expect both electron and hole sides to display a Fermi liquid phase by the exact same reasoning. Generally, we note that the system has no particle-hole symmetry with respect to $\nu=1$.

\emph{Numerical Analysis.--}
In order to support our hypothesis of a spin-polarized ground state, we have performed an exact diagonalization study of Eq.~\eqref{eq:single_orbital_hamiltonian} for finite systems with either periodic or twisted boundary conditions at filling factor $\nu=1$ for $N\in\lbrace 6,7,12,13,18,19\rbrace$ particles. The diagonalization is done using the matrix-free implicitly restarted Lanczos method.  We find the ground state at $\nu=1$ to be the fully spin-polarized state for all system sizes considered. Further technical details are given in the SM~\ref{sec:numerics}.

\begin{figure}[t]
    \centering
    \includegraphics[width=0.5\textwidth]{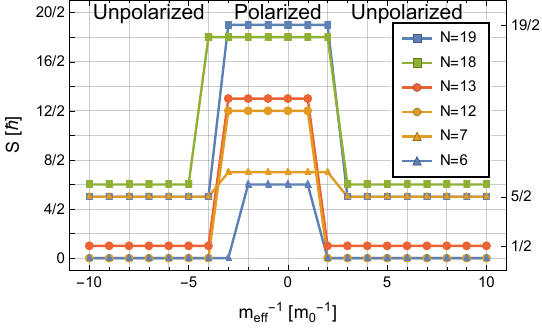}
    \caption{The ground state spin polarization of the Hamiltonian in Eq.~\eqref{eq:massive_hamiltonian} at $\nu=1$ as a function of inverse effective mass $m_{\text{eff}}^{-1}$, computed for various system sizes ranging from $N=6$ to $N=19$ particles. The horizontal axis increments are $m_{0}^{-1} = V_{k_0}N/(2Ak_0^2)$. The unpolarized phase has a sub-extensive polarization explained by Hund's rule.}
    \label{fig:spin_vs_m}
\end{figure}
To further solidify our claim, we have examined the robustness of the spin-polarized ground state to additional dispersion. We repeat the exact diagonalization at $\nu=1$ for the confined Coulomb Hamiltonian $\mathcal{H}_{\text{int}}$ in Eq.~\eqref{eq:single_orbital_hamiltonian} with an additional quadratic dispersion ($\hbar=1$):
\begin{equation}\label{eq:massive_hamiltonian}
    \mathcal{H}_{\text{int+mass}} = \mathcal{H}_{\text{int}} + \sum_{\sigma=\uparrow,\downarrow}\sum_{|\vec{k}|\leq k_0}
    \frac{|\vec{k}|^{2}}{2 m_{\text{eff}}}
    c_{\sigma,\vec{k}}^{\dagger}c_{\sigma,\vec{k}}.
\end{equation}
In Fig.~\ref{fig:spin_vs_m} we plot the total spin $S$ of the ground state of Eq.~\eqref{eq:massive_hamiltonian} at $\nu=1$ vs $m_{\text{eff}}^{-1}$, as the latter is swept from negative to positive values
in increments of $m_{0}^{-1} = V_{k_0}N/(2Ak_0^2)$. The flat kinetic dispersion in Eq.~\eqref{eq:well_dispersion} corresponds to $m_{\text{eff}}^{-1} = 0$. For the range of finite systems we considered, we find that the ground state is fully spin-polarized, i.e. $S=N/2$, for $-3m_0^{-1} \lesssim m_{\text{eff}}^{-1}\lesssim 2m_0^{-1}$, with the precise 
 phase boundaries slightly varying with system size. Outside the spin polarized region we find the spin polarization drops sharply to a lower value which depends on system size. One can show that, for all system sizes considered, this residual spin polarization has precisely the value expected by Hund's rule applied to a partial filling of the highest shell of degenerate kinetic energy states (see SM~\ref{sec:numerics}). We associate this result with a spin unpolarized phase, and expect that in the thermodynamic limit we will obtain $S/N\rightarrow 0$ in this regime of parameters.
The asymmetry of the spin-polarized regime boundary with respect to the sign of $m_{\text{eff}}^{-1}$ is due to the lower exchange energy of the unpolarized disk-shaped Fermi sea compared to that of an annular-shaped Fermi sea, the two different shapes that correspond to positive and negative $m_{\text{eff}}^{-1}$, respectively.  
In multilayer graphene systems with typical values of $k_0=0.2\textrm{nm}^{-1}$ and dielectric coefficient $\epsilon_{d}=6$, the mass scale is given by $m_{0}\sim\epsilon_{d} k_0/e^{2}\sim 0.06m_{e}$ where $m_{e}$ is the free electron mass.

\emph{Conclusions.--} We considered a model of interacting electrons confined in momentum space, and showed that when nearly half of the available states are filled, its ground state is spin polarized. We then numerically demonstrated that this spin polarization is stable to small variations of the model, and is not a fine-tuned consequence of the kinetic dispersion. We find this ground state to have a non-trivial excitation spectrum, including a diverging spin stiffness which leads to massless particle-hole excitations at low momenta. We argued in favor of a ferromagnetic Fermi liquid phase at $\nu=1\pm\varepsilon$, thereby demonstrating Stoner ferromagnetism in this model. The prevalence of spin and valley Stoner ferromagnets in multilayer rhombohedral graphene in a displacement field, where for a range of densities electrons are confined to a flat region of momentum space, may be understood within the framework of our model.

\let\oldaddcontentsline\addcontentsline
\renewcommand{\addcontentsline}[3]{}
\acknowledgements
OA would like to thank Johannes S. Hofmann, Daniel Kaplan, Tobias Holder and Shahar Barkai for the useful discussions. We acknowledge support from the Israeli Science Foundation Quantum Science and Technology grant no. 2074/19, and from the
CRC 183 of the Deutsche Forschungsgemeinschaft (Project C02). 
This work has received funding from the European Research Council (ERC) under the European Union’s Horizon 2020 research and innovation programme [grant agreements No. 788715 and 817799, Projects LEGOTOP (AS) and HQMAT (EB)].

\bibliography{main}
\let\addcontentsline\oldaddcontentsline

\newpage
\clearpage
\setcounter{figure}{0}
\setcounter{equation}{0}
\renewcommand{\theequation}{S\arabic{equation}}
\renewcommand{\thefigure}{S\arabic{figure}}
\renewcommand{\thesection}{S\arabic{section}}

\begin{widetext}
	
	\begin{center}
		\textbf{Supplemental Material}
    \end{center}

\tableofcontents

\section{Relation to rhombohedral graphite}\label{sec:rhombohedral_graphite}
In this section we show how the confining kinetic dispersion in Eq.~\eqref{eq:well_dispersion} can be thought of as a limit of infinitely many layers of rhombohedral graphite in the presence of a perpendicular electric field.
Consider a graphene multilayer whose sublattices at layer $n$ are denoted by $A_{n}$ and $B_{n}$. We say that the graphene is stacked rhombohedrally, or chirally\cite{Min2008-gu}, if the sublattice $B_{n+1}$ is directly above the sublattice $A_{n}$ for every $n$. For two layers this stacking configuration is known as Bernal stacking, while for three layers it is sometimes referred to as ABC stacking\cite{Zhang2010-xw}. In the derivation below we follow Ref.~\cite{Min2008-gu}, and assume the only interlayer tunneling is between $A_{n}$ and $B_{n+1}$. For $N_{l}$ layers, the effective low energy theory is hosted on sublattices $A_{1}$ and $B_{N_{l}}$, and is given (near one valley) by:
\begin{equation}
    H_{\text{eff}}(\vec{k}) = -t_{\perp}\begin{pmatrix}
    0 & \left(\frac{v_{D}}{t_\perp}(k_x-ik_y)\right)^{N_{l}} \\
    \left(\frac{v_{D}}{t_\perp}(k_x+ik_y)\right)^{N_{l}} & 0
    \end{pmatrix}.
\end{equation}
Here, $v_{D}$ is the monolayer Dirac velocity, and $t_{\perp}$ is the interlayer tunneling amplitude. This Hamiltonian displays a natural momentum scale given by $k_0=t_{\perp}/v_{D}$ which will simplify the notation below.

Next, consider applying a constant out-of-plane displacement field $D$. The two sites $A_{1}$ and $B_{N_{l}}$ are vertically separated by $(N_{l}-1)a_{\perp}$, where $a_{\perp}$ is the interlayer distance. The effective Hamiltonian in the presence of this displacement field can be written as:
\begin{equation}
    H_{\text{eff}}(\vec{k}) = \begin{pmatrix}
    \frac{eDa_{\perp}(N_{l}-1)}{2} & -t_{\perp}\left(\frac{k_x-ik_y}{k_0}\right)^{N_{l}} \\
    -t_{\perp}\left(\frac{k_x+ik_y}{k_0}\right)^{N_{l}} & -\frac{eDa_{\perp}(N_{l}-1)}{2}
    \end{pmatrix},
\end{equation}
which yields two bands, separated in energy by a gap that is roughly proportional to the displacement field and the number of layers:
\begin{equation}
    \epsilon_{\pm}(\vec{k}) = \pm \sqrt{
    \left(\frac{eDa_{\perp}(N_{l}-1)}{2}\right)^{2}
    +t_{\perp}^{2}\left(\frac{|\vec{k}|}{k_0}\right)^{2N_{l}}
    }.
\end{equation}
Considering many layers of graphene, i.e. $N_{l}\gg 1$, we find the following approximate energy dispersion:
\begin{equation}
    \epsilon_{\pm}(\vec{k}) \approx
    \begin{cases}
        \pm\frac{e\left|D\right|a_{\perp}N_{l}}{2}+\mathcal{O}\left(\left(\frac{|\vec{k}|}{k_0}\right)^{2N_{l}}\right), & |\vec{k}| < k_0 \\
        \pm\left|t_{\perp}\right|\left(\frac{|\vec{k}|}{k_0}\right)^{N_{l}}+\mathcal{O}\left(N_{l}^{2}\left(\frac{|\vec{k}|}{k_0}\right)^{-N_{l}}\right), & |\vec{k}| > k_0
    \end{cases}.
\end{equation}
In the limit of $N_{l}\rightarrow\infty$ this kinetic dispersion has a simple form,
\begin{equation}
    \epsilon_{+}(\vec{k})-\epsilon_{+}(0)=\begin{cases}
        0, & |\vec{k}| < k_0 \\
        \infty, & |\vec{k}| > k_0
    \end{cases},
\end{equation}
which is the confining kinetic dispersion in Eq.~\eqref{eq:well_dispersion}.

We note that, in practice, it is difficult to open a gap in rhombohedral graphite for too large $N_l$, since the perpendicular applied field is screened by the bulk electrons.

\section{The role of normal-ordering}\label{sec:normal_ordering}
In this section we derive the dispersion in Eq.~\eqref{eq:exact_h_dispersion} and calculate its explicit form. 
The momentum-confined Hamiltonian, as given by Eq.~\eqref{eq:single_orbital_hamiltonian}, is rewritten here as:
\begin{align}\label{eq:explicit_single_orbital_hamiltonian}
\begin{split}
    \mathcal{H}_{\text{int}} 
    = \frac{1}{2A}\sum_{\sigma,\sigma'=\uparrow,\downarrow}\sum_{\vec{q}}\sum_{\vec{k}\in D_{\vec{q}}}\sum_{\vec{k}'\in D_{-\vec{q}}}V_{\vec{q}}c_{\sigma',\vec{k}'-\vec{q}}^{\dagger}c_{\sigma,\vec{k}+\vec{q}}^{\dagger}c_{\sigma,\vec{k}}c_{\sigma',\vec{k}'}, \quad V_{\vec{q}}=2\pi e^{2}\frac{\tanh{(|\vec{q}|d)}}{|\vec{q}|},
\end{split}
\end{align}
and we reiterate our definition of $D_{\vec{q}}=\lbrace \vec{k}\enspace|\enspace\left(|\vec{k}|\leq k_0\right) \cap \left(|\vec{k}+\vec{q}|\leq k_0\right)\rbrace$, shown in Fiq.~\ref{fig:dq_region}.
By undoing the normal ordering, we find that:
\begin{align}
\begin{split}
    \mathcal{H}_{\text{int}} 
    &= \frac{1}{2A}\sum_{\sigma,\sigma'=\uparrow,\downarrow}\sum_{\vec{q}}\sum_{\vec{k}\in D_{\vec{q}}}\sum_{\vec{k}'\in D_{-\vec{q}}}V_{\vec{q}}c_{\sigma',\vec{k}'-\vec{q}}^{\dagger}c_{\sigma',\vec{k}'}c_{\sigma,\vec{k}+\vec{q}}^{\dagger}c_{\sigma,\vec{k}}
    -
    \frac{1}{2A}\sum_{\sigma,\sigma'=\uparrow,\downarrow}\sum_{\vec{q}}\sum_{\vec{k}\in D_{\vec{q}}}\sum_{\vec{k}'\in D_{-\vec{q}}}V_{\vec{q}}c_{\sigma',\vec{k}}^{\dagger}c_{\sigma,\vec{k}}\delta_{\vec{q},\vec{k}'-\vec{k}}\delta_{\sigma',\sigma}.
\end{split}
\end{align}
The first term is simply a density-density Hamiltonian for the projected particle density operator $\overline{\rho}_{\vec{q}}$ defined in Eq.~\eqref{eq:proj_particle_density}, and it is denoted in the main text by $\mathcal{H}_{\rho-\rho}$. The second term is the single particle dispersion we wish to calculate here. The dispersion contains a summation over momenta which is restricted such that $\left(|\vec{k}|\leq k_0\right) \cap \left(|\vec{k}+\vec{q}|\leq k_0\right)$ and $\left(|\vec{k}'|\leq k_0\right) \cap \left(|\vec{k}'-\vec{q}|\leq k_0\right)$.
The delta function $\delta_{\vec{q},\vec{k}'-\vec{k}}$ allows us to make the following change in the summation order:
\begin{equation}
    -\frac{1}{2A}\sum_{\sigma=\uparrow,\downarrow}\sum_{\vec{q}}\sum_{\vec{k}\in D_{\vec{q}}}\sum_{\vec{k}'\in D_{-\vec{q}}}V_{\vec{q}}c_{\sigma,\vec{k}}^{\dagger}c_{\sigma,\vec{k}}\delta_{\vec{q},\vec{k}'-\vec{k}} = -\frac{1}{2A}\sum_{\sigma=\uparrow,\downarrow}\sum_{\vec{k}\in D_{0}}\sum_{\vec{k}'\in D_{0}}\sum_{\vec{q}}V_{\vec{q}}c_{\sigma,\vec{k}}^{\dagger}c_{\sigma,\vec{k}}\delta_{\vec{q},\vec{k}'-\vec{k}}.
\end{equation}
Performing the summation over $\vec{q}$, we find the dispersion:
\begin{equation}
    -
    \frac{1}{2A}\sum_{\sigma=\uparrow,\downarrow}\sum_{\vec{k}\in D_{0}}\sum_{\vec{k}'\in D_{0}}\sum_{\vec{q}}V_{\vec{q}}c_{\sigma,\vec{k}}^{\dagger}c_{\sigma,\vec{k}}\delta_{\vec{q},\vec{k}'-\vec{k}}
    =
    -
    \sum_{\sigma=\uparrow,\downarrow}
    \sum_{\vec{k}\in D_{0}}E_{0}(\vec{k})c_{\sigma,\vec{k}}^{\dagger}c_{\sigma,\vec{k}},
    \quad
    E_{0}(\vec{k})=\frac{1}{2A}\sum_{|\vec{k}'|\leq k_0}V_{\vec{k}-\vec{k}'},
\end{equation}
which is the result presented in Eq.~\eqref{eq:exact_h_dispersion}.

We now proceed to explicitly calculate the dispersion $E_{0}(\vec{k})$ in the thermodynamic limit. Rewriting the sum as an integral, we find:
\begin{align}
\begin{split}
    E_{0}(\vec{k}) =  \frac{1}{2A}\sum_{|\vec{k}'|\leq k_0}V_{\vec{k}-\vec{k}'} &= \frac{1}{2}\int_{|\vec{k}'|\leq k_0}\frac{d^2 k'}{(2\pi)^{2}}V_{\vec{k}-\vec{k}'}\\
    &=\frac{1}{2}\int_{|\vec{p}-\vec{k}|\leq k_0}\frac{d^{2}p}{(2\pi)^{2}}V_{\vec{p}}
    \\
    &=\pi e^{2} \int_{|\vec{p}-\vec{k}|\leq k_0}\frac{d^{2}p}{(2\pi)^{2}}\frac{\tanh{(|\vec{p}|d)}}{|\vec{p}|}
    \\
    &=\pi e^{2} \int_{-\pi}^{\pi} \frac{d\theta}{2\pi}\int_{0}^{\sqrt{k_{0}^{2}-|\vec{k}|^{2}\sin^{2}{\theta}}+|\vec{k}|\cos{\theta}}\frac{pdp}{2\pi}\frac{\tanh(pd)}{p},
\end{split}
\end{align}
and by performing the radial integration we obtain the result:
\begin{equation}\label{eq:integral_form_of_e0}
    E_{0}(k) = \frac{e^{2}k_0}{2 k_0 d}\int_{-\pi}^{\pi} \frac{d\theta}{2\pi}\ln\cosh\left[\left(\sqrt{k_{0}^{2}-k^{2}\sin^{2}{\theta}}+k\cos{\theta}\right)d\right],
\end{equation}
where we denoted $|\vec{k}|=k$, as the dispersion has rotational symmetry. 

The result in Eq.~\eqref{eq:integral_form_of_e0} is plotted in Fig.~\ref{fig:e0_vs_k} in the main text. This integral has a familiar form in the limit of unscreened interactions $k_0d\rightarrow\infty$. We now present it along with an expansion in powers of $(k_0d)^{-1}$.
Before continuing, it is useful to switch to dimensionless variables. 
Denote $f(x)=E_{0}(xk_0)/(e^{2}k_0)$ and $g(x,\theta) = \sqrt{1-x^{2}\sin^{2}{\theta}}+x\cos{\theta}$ such that:
\begin{equation}
    f(x) = \frac{1}{2 k_0 d}\int_{-\pi}^{\pi} \frac{d\theta}{2\pi}\ln\cosh\left[g(x,\theta)k_0 d\right].
\end{equation}
Evaluation of the logarithm reveals the leading order terms in $(k_0d)^{-1}$:
\begin{align}\label{eq:screening_expansion_of_dispersion}
\begin{split}
    f(x) &= \frac{1}{2 k_0 d}\int_{-\pi}^{\pi} \frac{d\theta}{2\pi}\left(-\ln{2}+g(x,\theta)k_0 d + \ln{\left(1+e^{-2g(x,\theta)k_0d}\right)}\right) \\
    &= \frac{1}{2}\int_{-\pi}^{\pi} \frac{d\theta}{2\pi}\sqrt{1-x^{2}\sin^{2}{\theta}}
    +\frac{1}{2k_0 d}\left(-\ln{2}+\int_{-\pi}^{\pi} \frac{d\theta}{2\pi}\ln{\left(1+e^{-2g(x,\theta)k_0d}\right)}\right) \\
    &= \frac{1}{\pi}\tilde{E}\left(\frac{\pi}{2},x\right)
    +\frac{1}{2k_0 d}\left(-\ln{2}+\int_{-\pi}^{\pi} \frac{d\theta}{2\pi}\ln{\left(1+e^{-2g(x,\theta)k_0d}\right)}\right), \\
\end{split}
\end{align}
where $\tilde{E}\left(\frac{\pi}{2},x\right)$ is the complete elliptic integral of the second kind. 

The first term in Eq.~\eqref{eq:screening_expansion_of_dispersion} is a closed form solution for $f(x)$ to zeroth order in $(k_0d)^{-1}$. 
For the next order in $(k_0d)^{-1}$, we need to evaluate the integral in the third term and determine its behavior in the limit of $k_0d\rightarrow\infty$. To do so, we separate our discussion into two regimes: $x<1$ and $x=1$. For $x<1$ we have $g(x,\theta)>0$ for all $\theta$, while for $x=1$, however, $g(x,\theta)>0$ only for $\frac{\pi}{2}<|\theta|<\pi$, otherwise it vanishes. This implies the following limits:
\begin{equation}
    \lim_{k_0d\rightarrow\infty}\int_{-\pi}^{\pi} \frac{d\theta}{2\pi}\ln{\left(1+e^{-2g(x,\theta)k_0d}\right)}
    = 
    \begin{cases}
        0, & x< 1,\\
        \frac{\ln{2}}{2}, & x=1.
    \end{cases}
\end{equation}
We conclude that the power series for $f(x)$ does not converge uniformly:
\begin{equation}
    f(x)
    \approx 
    \begin{cases}
        \frac{1}{\pi}\tilde{E}\left(\frac{\pi}{2},x\right)
        -\frac{\ln{2}}{2 k_0 d} +\mathcal{O}\left((k_0d)^{-2}\right), & x< 1,\\
        \frac{1}{\pi}\tilde{E}\left(\frac{\pi}{2},x\right)
        -\frac{\ln{2}}{4k_0 d}+\mathcal{O}\left((k_0d)^{-2}\right), & x=1.
    \end{cases}
\end{equation}
Re-introducing the physical units of $e^2$ and $k_0$, we find the dispersion to leading order in $(k_0d)^{-1}$ to be:
\begin{equation}\label{eq:leading_order_k0d}
    E_{0}(\vec{k}) =
    \begin{cases}
        e^{2}k_0\left(\frac{1}{\pi}\tilde{E}\left(\frac{\pi}{2},\frac{|\vec{k}|}{k_0}\right)-\frac{1}{2}\frac{\ln{2}}{k_0 d}\right), & |\vec{k}|<k_0, \\ \\
        e^{2}k_0\left(\frac{1}{\pi}\tilde{E}\left(\frac{\pi}{2},1\right)-\frac{1}{4}\frac{\ln{2}}{k_0 d}\right), & |\vec{k}|=k_0,
    \end{cases}
\end{equation}
and in the limit of $k_0d\rightarrow\infty$ we obtain the simple form presented in Eq.~\eqref{eq:unscreened_h_dispersion}. The seeming discontinuity in Eq.~\eqref{eq:leading_order_k0d} comes from the divergence of the derivative of $\tilde{E}\left(\frac{\pi}{2},\frac{|\vec{k}|}{k_0}\right)$ at $|\vec{k}|=k_0$.

\section{Collective particle-hole excitations}\label{sec:collective_ph_excitations}
In this section we examine collective particle-hole excitations with momentum $\vec{Q}$ above the spin-polarized state, which by definition are magnons. We consider both the strictly confining kinetic dispersion and a softened version of the momentum confinement. For both cases we are interested in the dispersion relation of the lowest-lying magnon excitation. We show how the former results in a linear dispersion (and calculate its velocity), while for the latter we find the usual quadratic magnon dispersion and calculate its spin stiffness.

\subsection{The strictly confining kinetic dispersion}
\subsubsection{Constructing the collective excitation matrix}
The collective particle-hole excitations are eigenstates of $\mathcal{H}_{\text{int}}$ which are spanned by single particle-hole excited states denoted by $\ket{\Psi_{\text{ph}}^{\vec{k},\vec{Q}}}$ and defined in Eq.~\eqref{eq:col_exc_basis}. In what follows we derive the matrix in Eq.~\eqref{eq:col_exc_mat_inf} whose eigenstates are the collective excitations. We therefore consider the following matrix element:
\begin{align}\label{eq:collective_matrix_element}
\begin{split}
    [H(\vec{Q})]_{\vec{k},\vec{k}'} &=\bra{\Psi_{\text{ph}}^{\vec{k}',\vec{Q}}}\mathcal{H}_{\text{int}}\ket{\Psi_{\text{ph}}^{\vec{k},\vec{Q}}} \\
    &= \bra{\Psi_{\text{SP}}}c_{\downarrow,\vec{k}'}^{\dagger}c_{\uparrow,\vec{k}'+\vec{Q}}\mathcal{H}_{\text{int}}c_{\uparrow,\vec{k}+\vec{Q}}^{\dagger}c_{\downarrow,\vec{k}}\ket{\Psi_{\text{SP}}},
\end{split}
\end{align}
for $\vec{k},\vec{k}'\in D_{\vec{Q}}$, and $D_{\vec{Q}}=\lbrace \vec{k} \enspace | \enspace(|\vec{k}|\leq k_{0}) \cap (|\vec{k}+\vec{Q}|\leq k_{0})\rbrace$. The wavefunction $\ket{\Psi_{\text{SP}}}=\prod_{\vec{k}\in D_{0}}c_{\downarrow,\vec{k}}^{\dagger}\ket{0}$ is the one described in the main text, and is obtained by completely filling the disk $D_{0}$ with spin down electrons. The Hamiltonian $\mathcal{H}_{\text{int}}$ is defined in Eq.~\eqref{eq:single_orbital_hamiltonian}, and is written more explicitly in Eq.~\eqref{eq:explicit_single_orbital_hamiltonian}. 
From the explicit form of $\mathcal{H}_{\text{int}}$ we see that the matrix element $[H(\vec{Q})]_{\vec{k},\vec{k}'}$ in Eq.~\eqref{eq:collective_matrix_element} is a sum over contractions of the form:
\begin{align}\label{eq:contractions}
    \begin{split}
    &\bra{\Psi_{\text{SP}}}c_{\downarrow,\vec{k}'}^{\dagger}c_{\uparrow,\vec{k}'+\vec{Q}}\left(c_{\sigma,\vec{p}+\vec{q}}^{\dagger}c_{\sigma',\vec{p}'-\vec{q}}^{\dagger}c_{\sigma',\vec{p}'}c_{\sigma,\vec{p}}\right)c_{\uparrow,\vec{k}+\vec{Q}}^{\dagger}c_{\downarrow,\vec{k}}\ket{\Psi_{\text{SP}}}\\
    &=
    1_{\vec{k}\in D_{0}}\left[\delta_{\sigma,\downarrow}\delta_{\sigma',\downarrow}\delta_{\vec{k},\vec{k}'}1_{\vec{p}\in D_{0}}1_{\vec{p}'\in D_{0}}(\delta_{\vec{q},0}-\delta_{\vec{q},\vec{p}'-\vec{p}})\right.
    \\
    &+
    \delta_{\vec{k},\vec{k}'}\delta_{\vec{q},0}\left(
    \delta_{\sigma,\downarrow}
    \left(
    \delta_{\sigma',\uparrow}\delta_{\vec{p}',\vec{k}+\vec{Q}}
    -\delta_{\sigma',\downarrow}\delta_{\vec{p}',\vec{k}}
    \right)
    1_{\vec{p}\in D_{0}}
    +
    \delta_{\sigma',\downarrow}
    \left(
    \delta_{\sigma,\uparrow}\delta_{\vec{p},\vec{k}+\vec{Q}}
    -\delta_{\sigma,\downarrow}\delta_{\vec{k},\vec{p}}
    \right)
    1_{\vec{p}'\in D_{0}}\right)
    \\
    &+\delta_{\vec{k},\vec{k}'}
    \left(
    \delta_{\sigma,\downarrow}\delta_{\sigma',\downarrow}\delta_{\vec{q},\vec{p}'-\vec{k}}
    \delta_{\vec{k},\vec{p}}
    +\delta_{\sigma,\downarrow}\delta_{\sigma',\downarrow}\delta_{\vec{q},\vec{k}-\vec{p}}
    \delta_{\vec{k},\vec{p}'}
    \right)1_{\vec{p}\in D_{0}}1_{\vec{p}'\in D_{0}}
    \\
    &\left.
    -
    \delta_{\sigma,\uparrow}\delta_{\sigma',\downarrow}\delta_{\vec{p},\vec{k}+\vec{Q}}\delta_{\vec{q},\vec{k}'-\vec{k}}
    \delta_{\vec{k}',\vec{p}'}
    1_{\vec{p}'\in D_{0}}
    -\delta_{\sigma,\downarrow}\delta_{\sigma',\uparrow}\delta_{\vec{p}',\vec{k}+\vec{Q}}\delta_{\vec{q},\vec{k}-\vec{k}'}
    \delta_{\vec{k}',\vec{p}}
    1_{\vec{p}\in D_{0}}\right].
    \end{split} 
\end{align}
The matrix element is calculated by carefully performing the summation over Eq.~\eqref{eq:contractions} as constrained by Eq.~\eqref{eq:explicit_single_orbital_hamiltonian} for $\vec{k},\vec{k}'\in D_{\vec{Q}}$:
\begin{equation}
    \bra{\Psi_{\text{ph}}^{\vec{k}',\vec{Q}}}\mathcal{H}_{\text{int}}\ket{\Psi_{\text{ph}}^{\vec{k},\vec{Q}}}
    =
    \delta_{\vec{k},\vec{k}'}
    \left(
    E_{\text{SP}}
    +\frac{1}{A}\sum_{\vec{p}\in D_{0}}V_{\vec{k}-\vec{p}}
    \right)
    -\frac{1}{A}V_{\vec{k}-\vec{k}'},
\end{equation}
where we identified the spin polarized energy $E_{\text{SP}}=\braket{\Psi_{\text{SP}}|\mathcal{H}_{\text{int}}|\Psi_{\text{SP}}}=\frac{1}{2A}\sum_{\vec{p},\vec{p}'\in D_{0}}\left(V_{0}-V_{\vec{p}-\vec{p}'}\right)$.

The collective excitation matrix is confined to $\vec{k},\vec{k}'\in D_{\vec{Q}}$ and is given by:
\begin{equation}\label{eq:collective_matrix}
    [H(\vec{Q})]_{\vec{k},\vec{k}'} = 
        \delta_{\vec{k},\vec{k}'}
        \left(
        E_{\text{SP}}
        +\frac{1}{A}\sum_{\vec{p}\in D_{0}}V_{\vec{k}-\vec{p}}
        \right)
        -\frac{1}{A}V_{\vec{k}-\vec{k}'}, 
         \quad
        \vec{k},\vec{k}'\in D_{\vec{Q}}.
\end{equation}

\subsubsection{Excitation spectrum bounds}
Diagonalizing the matrix in Eq.~\eqref{eq:collective_matrix} can be done numerically for any $\vec{Q}$, and the results are given in Fig.~\ref{fig:ph_spectrum}. Analytically, we can derive bounds on parts of the excitation spectrum which will be useful for the analytical structure of the lowest-lying eigenvalue in the vicinity of $\vec{Q}=0$. We demonstrate the existence of a gapless particle-hole excitation mode, which is expected, and prove that there is precisely one such mode.

We are interested in the excitation energies of Eq.~\eqref{eq:collective_matrix} with respect to the ground state energy. We therefore shift our reference energy by defining:
\begin{equation}\label{eq:collective_matrix_shifted}
    \tilde{H}(\vec{Q}) = H(\vec{Q})-E_{SP},
\end{equation}
and denote its corresponding eigensystem ($n=1,\dots ,|D_{\vec{Q}}|$) by:
\begin{equation}\label{eq:ph_eigensystem}
    \tilde{H}(\vec{Q})\ket{\Psi_{ph}^{(n)}(\vec{Q})}=E_{\text{ph},n}(\vec{Q})\ket{\Psi_{ph}^{(n)}(\vec{Q})},\quad E_{\text{ph},1}(\vec{Q})\leq E_{\text{ph},2}(\vec{Q}) \leq \dots \leq E_{\text{ph},|D_{\vec{Q}}|}(\vec{Q}).
\end{equation}
In what follows we will find upper and lower bounds on the excitation energies $E_{\text{ph},n}(\vec{Q})$, and show that the only excitation mode that has zero energy is at $n=1$ and $\vec{Q}=0$.

The matrix $\tilde{H}(\vec{Q})$ in Eq.~\eqref{eq:collective_matrix_shifted} is a Hermitian diagonally dominant matrix with positive diagonal entries, as can be easily verified from Eq.~\eqref{eq:collective_matrix}. Using the Gershgorin circle theorem\cite{Gershgorin1931-ss,Varga2011-ha}, it is easy to show that the eigenvalues of $\tilde{H}(\vec{Q})$, $E_{\text{ph},n}(\vec{Q})$, are bounded by:
\begin{equation}\label{eq:gershgorin_bound}
    \min_{\vec{k}\in D_{\vec{Q}}}\lbrace\frac{1}{A}\sum_{\vec{k}'\in D_{0}}V_{\vec{k}-\vec{k}'}-\frac{1}{A}\sum_{\substack{\vec{k}'\in D_{\vec{Q}}\\\vec{k}'\neq\vec{k}}}V_{\vec{k}-\vec{k}'}\rbrace \leq E_{\text{ph},n}(\vec{Q}) \leq \max_{\vec{k}\in D_{\vec{Q}}}\lbrace\frac{1}{A}\sum_{\vec{k}'\in D_{0}}V_{\vec{k}-\vec{k}'}+\frac{1}{A}\sum_{\substack{\vec{k}'\in D_{\vec{Q}}\\\vec{k}'\neq\vec{k}}}V_{\vec{k}-\vec{k}'}\rbrace.
\end{equation}
For $\vec{Q}\neq 0$, this implies that $E_{\text{ph},n}(\vec{Q})>0$ are strictly positive. For $\vec{Q}=0$, however, we find the following bounds (in the thermodynamic limit):
\begin{equation}\label{eq:upper_bound_on_ph_energy}
    0 \leq E_{\text{ph},n}(\vec{Q}=0) \leq 2\max_{\vec{k}}\lbrace\frac{1}{A}\sum_{\vec{k}'\in D_{0}}V_{\vec{k}-\vec{k}'}\rbrace = \left. 4E_{0}(\vec{k})\right|_{\vec{k}=0} \leq 2e^{2}k_0,
\end{equation}
where $E_{0}(\vec{k})$ is defined in Eq.~\eqref{eq:exact_h_dispersion}. 

The upper bound in Eq.~\eqref{eq:upper_bound_on_ph_energy} will suffice for our purposes. The lower bound, however, will be improved momentarily. Before we do so, we make a detour to simplify notation. We denote the basis vectors $\ket{\Psi_{\text{ph}}^{\vec{k},\vec{Q}}}$ by $\ket{\vec{k}}$ where the momentum $\vec{Q}$ is implied by context, such that:
\begin{equation}\label{eq:simplified_ph_basis_notation}
    \tilde{H}(\vec{Q})=\sum_{\vec{k},\vec{k}'\in D_{\vec{Q}}}[\tilde{H}(\vec{Q})]_{\vec{k},\vec{k}'}\ket{\vec{k}}\bra{\vec{k}'}.
\end{equation}
For $\vec{Q}=0$, there are $N=|D_{0}|$ such basis vectors. Returning to the effort to tighten the lower bound, consider a normalized eigenvector $\ket{\Psi_{ph}^{(n)}(0)}$ of $\tilde{H}(\vec{Q}=0)$. Its eigenvalue $E_{\text{ph},n}(0)$ is given by:
\begin{align}\label{eq:lower_bound_derivation_q0}
\begin{split}
    E_{\text{ph},n}(0) &= \braket{\Psi_{ph}^{(n)}(0)|\tilde{H}(\vec{Q}=0)|\Psi_{ph}^{(n)}(0)}
    \\
    &= \sum_{\vec{k},\vec{k}'\in D_{0}}[\tilde{H}(\vec{Q}=0)]_{\vec{k},\vec{k}'}\braket{\Psi_{ph}^{(n)}(0)|\vec{k}}\braket{\vec{k}'|\Psi_{ph}^{(n)}(0)}
    \\
    &= \sum_{\vec{k}\in D_{0}}[\tilde{H}(\vec{Q}=0)]_{\vec{k},\vec{k}}\braket{\Psi_{ph}^{(n)}(0)|\vec{k}}\braket{\vec{k}|\Psi_{ph}^{(n)}(0)}
    +\sum_{\substack{\vec{k},\vec{k}'\in D_{0}\\\vec{k}\neq\vec{k}'}}[\tilde{H}(\vec{Q}=0)]_{\vec{k},\vec{k}'}\braket{\Psi_{ph}^{(n)}(0)|\vec{k}}\braket{\vec{k}'|\Psi_{ph}^{(n)}(0)}
    \\
    &= \sum_{\substack{\vec{k},\vec{k}'\in D_{0}\\\vec{k}\neq\vec{k}'}}[\tilde{H}(\vec{Q}=0)]_{\vec{k},\vec{k}'}\left(\braket{\Psi_{ph}^{(n)}(0)|\vec{k}}\braket{\vec{k}'|\Psi_{ph}^{(n)}(0)}-\braket{\Psi_{ph}^{(n)}(0)|\vec{k}}\braket{\vec{k}|\Psi_{ph}^{(n)}(0)}\right)
    \\
    &= \frac{1}{2}\sum_{\substack{\vec{k},\vec{k}'\in D_{0}\\\vec{k}\neq\vec{k}'}}\left(-[\tilde{H}(\vec{Q}=0)]_{\vec{k},\vec{k}'}\right)\left|\braket{\vec{k}'|\Psi_{ph}^{(n)}(0)}-\braket{\vec{k}|\Psi_{ph}^{(n)}(0)}\right|^{2} 
    \\
    &\geq \frac{1}{2}\min_{\substack{\vec{k},\vec{k}'\in D_{0}\\\vec{k}\neq\vec{k}'}}\lbrace-[\tilde{H}(\vec{Q}=0)]_{\vec{k},\vec{k}'}\rbrace\sum_{\vec{k},\vec{k}'\in D_{0}}\left|\braket{\vec{k}'|\Psi_{ph}^{(n)}(0)}-\braket{\vec{k}|\Psi_{ph}^{(n)}(0)}\right|^{2},
\end{split}
\end{align}
where we used the fact that the diagonal is given by the sum of the rest of the row, the symmetry of the matrix, and that its off-diagonal elements are negative. The minimal off-diagonal element is given by:
\begin{equation}\label{eq:lower_bound_minimal_element_q0}
    \min_{\substack{\vec{k},\vec{k}'\in D_{0}\\\vec{k}\neq\vec{k}'}}\lbrace-[\tilde{H}(\vec{Q}=0)]_{\vec{k},\vec{k}'}\rbrace = \frac{V_{2k_0}}{A},
\end{equation}
which implies:
\begin{equation}\label{eq:lower_bound_q0}
    E_{\text{ph},n}(0) \ge \frac{V_{2k_0}}{2A}\sum_{\vec{k},\vec{k}'\in D_{0}}\left|\braket{\vec{k}'|\Psi_{ph}^{(n)}(0)}-\braket{\vec{k}|\Psi_{ph}^{(n)}(0)}\right|^{2}.
\end{equation}

The lower bound in Eq.~\eqref {eq:lower_bound_q0} can be computed exactly and therefore compared to the lower bound found in Eq.~\eqref{eq:upper_bound_on_ph_energy}. For $n=1$ the two lower bounds coincide, as they are satisfied by the uniform vector $\ket{\Psi_{ph}^{(1)}(\vec{Q}=0)}=\frac{1}{\sqrt{N}}\sum_{\vec{k}\in D_{0}}\ket{\vec{k}}$, i.e. $\tilde{H}(\vec{Q}=0)\ket{\Psi_{ph}^{(1)}(0)}=E_{\text{ph},1}(0)\ket{\Psi_{ph}^{(1)}(\vec{Q}=0)}=0$. For $n>1$, the bound in Eq.~\eqref {eq:lower_bound_q0} is stricter than the bound in Eq.~\eqref{eq:upper_bound_on_ph_energy}. To show that, we make use of the orthogonality of eigenvectors of $\tilde{H}(\vec{Q}=0)$:
\begin{equation}\label{eq:orth_conseq}
    n>1: \quad \braket{\Psi_{ph}^{(1)}(0)|\Psi_{ph}^{(n)}(0)} = 0 \implies \sum_{\vec{k}\in D_{0}}\braket{\vec{k}|\Psi_{ph}^{(n)}(0)} = 0,
\end{equation}
and use the above result to calculate the sum in Eq.~\eqref {eq:lower_bound_q0} exactly for $n>1$:
\begin{equation}\label{eq:lower_bound_wavefunction_q0}
    \sum_{\vec{k},\vec{k}'\in D_{0}}\left|\braket{\vec{k}'|\Psi_{ph}^{(n)}(0)}-\braket{\vec{k}|\Psi_{ph}^{(n)}(0)}\right|^{2} = 2\left(\sum_{\vec{k},\vec{k}'\in D_{0}}\left|\braket{\vec{k}|\Psi_{ph}^{(n)}(0)}\right|^{2} - \sum_{\vec{k}'\in D_{0}}\braket{\Psi_{ph}^{(n)}(0)|\vec{k}'}\sum_{\vec{k}\in D_{0}}\braket{\vec{k}|\Psi_{ph}^{(n)}(0)}\right) = 2N,
\end{equation}
where we used the normalization of $\ket{\Psi_{ph}^{(n)}(0)}$, $|D_{0}|=N$, and Eq.~\eqref{eq:orth_conseq}. Inserting this result back into Eq.~\eqref {eq:lower_bound_q0}, we find the lower bound on the gap to the next excitation energy, $E_{\text{ph},n}(0) \geq V_{2k_0}(N/A)$. Taking the thermodynamic limit, at filling $\nu=1$ we have $N/A=k_{0}^{2}/(4\pi)$, which implies $E_{\text{ph},n}(0) \geq e^{2}k_{0}/4$. Combining this result with the upper bound in Eq.~\eqref{eq:upper_bound_on_ph_energy}, we find:
\begin{equation}\label{eq:bound_on_ph_energy}
    \begin{cases}
        E_{\text{ph},1}(\vec{Q}=0) = 0, & n = 1, \\
        \frac{1}{4}e^{2}k_{0} \leq E_{\text{ph},n}(\vec{Q}=0) \leq 2e^{2}k_{0}, & n> 1.
    \end{cases}
\end{equation}
We conclude there is precisely one excitation energy that starts at zero energy, and the rest are gapped excitations.

\begin{figure}
    \centering
    \includegraphics[width=\textwidth]{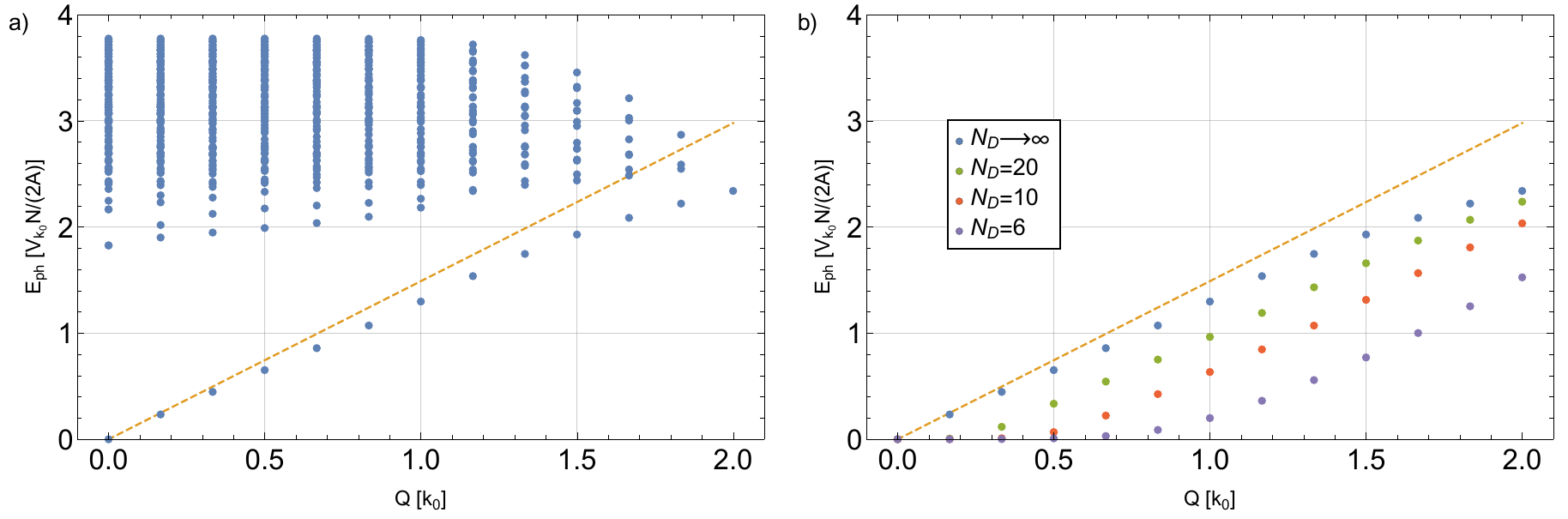}
    \caption{Particle-hole excitation spectrum for $k_0d\rightarrow\infty$. In figure (a) we plot the entire spectrum for the strict momentum confinement. The lowest branch begins with a linear slope. The dashed line is linear whose slope is the one theoretically predicted (computed for finite system size $N$). Note that the number of finite excitations decreases with increasing $Q$ as the domain of our Hamiltonian becomes smaller.
    In figure (b) we compare the lowest branch of the strict momentum confinement (and its linear slope) with that of the softened confinement of different powers $N_{D}$ and the same energy scale $U_{\text{K}}=0.1V_{k_0}N/A$. The crossover to linear slope increases with $N_{D}$, and in the limit $N_{D}\rightarrow \infty$ it reaches the origin.}
    \label{fig:ph_spectrum}
\end{figure}

\subsubsection{Dispersion of the lowest excitation energy}
We are interested in calculating the evolution of the lowest excitation energy with $\vec{Q}$. We will show here that it is linear in $|\vec{Q}|$, and calculate its velocity as shown in Eq.~\eqref{eq:linear_ph_disp}. We will do so by bounding the energy from both sides by two bounds with the same linear behavior at small $|\vec{Q}|$.

Our starting point is the collective excitation matrix as defined in Eq.~\eqref{eq:collective_matrix_shifted}:
\begin{align}
\begin{split}
    \tilde{H}(\vec{Q}) 
    &= \sum_{\vec{k}\in D_{\vec{Q}}}
        2E_{0}(\vec{k})\ket{\vec{k}}\bra{\vec{k}}
        -
        \sum_{\vec{k},\vec{k}'\in D_{\vec{Q}}}
        \frac{1}{A}V_{\vec{k}-\vec{k}'}
        \ket{\vec{k}}\bra{\vec{k}'}.
\end{split}
\end{align}
We decompose it as $\tilde{H}(\vec{Q})=\tilde{H}_{0}(\vec{Q})+\tilde{H}_{1}(\vec{Q})$ such that:
\begin{align}
    \tilde{H}_{0}(\vec{Q})&=
    \sum_{\vec{k}\in D_{\vec{Q}}}
    \left(
    \frac{1}{A}\sum_{\vec{p}\in D_{\vec{Q}}}V_{\vec{k}-\vec{p}}
    \right)\ket{\vec{k}}\bra{\vec{k}}
    -
    \sum_{\vec{k},\vec{k}'\in D_{\vec{Q}}}
    \frac{1}{A}V_{\vec{k}-\vec{k}'}
    \ket{\vec{k}}\bra{\vec{k}'},
    \\
    \tilde{H}_{1}(\vec{Q})&=
    \sum_{\vec{k}\in D_{\vec{Q}}}
    \left(
    \frac{1}{A}\sum_{\vec{p}\in D_{0}-D_{\vec{Q}}}V_{\vec{k}-\vec{p}}
    \right)\ket{\vec{k}}\bra{\vec{k}}.
\end{align}
The purpose of this decomposition will be clear momentarily. The diagonal elements of $\tilde{H}_{1}(\vec{Q})$ can be bounded as follows:
\begin{align}\label{eq:exchange_of_lune}
\begin{split}
    \frac{1}{A}\sum_{\vec{p}\in D_{0}-D_{\vec{Q}}}V_{\vec{k}-\vec{p}}
    &= \int_{\vec{p}\in D_{0}-D_{\vec{Q}}}\frac{d^{2}p}{(2\pi)^{2}}V_{\vec{k}-\vec{p}} \\
    &\leq 
    \int_{-\frac{|\vec{Q}|}{2}}^{\frac{|\vec{Q}|}{2}}\frac{dq_{x}}{2\pi}
    \int_{-k_0}^{k_0}\frac{dq_{y}}{2\pi}V_{\vec{q}}\\
    &\leq \frac{2}{\pi}e^{2}
    \int_{0}^{\frac{|\vec{Q}|}{2}}dq_{x}
    \int_{0}^{k_0}\frac{dq_{y}}{\sqrt{q_{x}^2+q_{y}^2}}\\
    &= \frac{2}{\pi}e^{2}
    \int_{0}^{\frac{|\vec{Q}|}{2}}dq_{x}
    \sinh^{-1}\left(\frac{k_0}{q_{x}}\right)\\
    &= \frac{1}{\pi}e^{2}\frac{|\vec{Q}|}{k_0}\log{\left(\frac{k_0}{|\vec{Q}|}\right)} +\mathcal{O}\left(\frac{|\vec{Q}|}{k_0}\right).
\end{split}
\end{align}
In the bound above, the first inequality is a result of slicing the region $D_{0}-D_{\vec{Q}}$ into slabs of length $|\vec{Q}|$ along $q_{x}$ (i.e. $\hat{Q}$) and width $dq_{y}$, and shifting them along $q_{x}$ to maximize the integrand. The second inequality is simply bounding the (possibly) screened Coulomb potential by the unscreened potential with $k_0d\rightarrow\infty$.

Next, consider a uniform vector $\ket{\phi_{0}(\vec{Q})}=\frac{1}{\sqrt{|D_{\vec{Q}}|}}\sum_{\vec{k}\in D_{\vec{Q}}}\ket{\vec{k}}$. One can easily verify that $\tilde{H}_{0}(\vec{Q})\ket{\phi_{0}(\vec{Q})}=0$. The expectation value for the energy in this state is given by:
\begin{align}
\begin{split}
    \bra{\phi_{0}(\vec{Q})}\tilde{H}(\vec{Q})\ket{\phi_{0}(\vec{Q})} 
    &= \bra{\phi_{0}(\vec{Q})}\tilde{H}_{1}(\vec{Q})\ket{\phi_{0}(\vec{Q})}
    \\
    &= \frac{2}{|D_{\vec{Q}}|}\sum_{\vec{k}\in D_{\vec{Q}}}
    \left(
    \frac{1}{2A}
    \sum_{\vec{k}'\in D_{0}-D_{\vec{Q}}}
    V_{\vec{k}-\vec{k}'}
    \right)
    \\
    &= \frac{2}{|D_{\vec{Q}}|}
    \sum_{\vec{k}'\in D_{0}-D_{\vec{Q}}}
    \left(
    E_0(\vec{k}')
    -
    \frac{1}{2A}
    \sum_{\vec{k}\in D_{0}-D_{\vec{Q}}}
    V_{\vec{k}-\vec{k}'}
    \right),
\end{split}
\end{align}
Keeping only the linear order in $|\vec{Q}|/k_0$, and making use of the bound in Eq.~\eqref{eq:exchange_of_lune}, we find:
\begin{align}\label{eq:uniform_expectation_vs_q}
\begin{split}
    \bra{\phi_{0}(\vec{Q})}\tilde{H}(\vec{Q})\ket{\phi_{0}(\vec{Q})} 
    &\approx 2\frac{|D_0-D_{\vec{Q}}|}{|D_{0}|}E_{0}(k_0)
    +\mathcal{O}\left(\frac{|\vec{Q}|^{2}}{k_0^{2}}\log{\left(\frac{|\vec{Q}|}{k_0}\right)}\right)
    \\
    &\approx 2\frac{2 k_0 |\vec{Q}|}{\pi k_0^{2}}E_{0}(k_0)
    +\mathcal{O}\left(\frac{|\vec{Q}|^{2}}{k_0^{2}}\log{\left(\frac{|\vec{Q}|}{k_0}\right)}\right).
\end{split}
\end{align}
By definition, the lowest eigenvalue is bounded from above by the expectation value of $\tilde{H}(\vec{Q})$ with respect to any normalized wavefunction. Therefore, we find that:
\begin{equation}\label{eq:upper_bound_on_dispersion_vs_Q}
    E_{\text{ph},1}(\vec{Q})\leq \frac{4 E_{0}(k_0)}{\pi k_0} |\vec{Q}| +\mathcal{O}\left(\frac{|\vec{Q}|^{2}}{k_0^{2}}\log{\left(\frac{|\vec{Q}|}{k_0}\right)}\right)
\end{equation}

We now show that the lower bound is equal to the upper bound in Eq.~\eqref{eq:upper_bound_on_dispersion_vs_Q}. Consider $\ket{\Psi_{ph}^{(1)}(\vec{Q})}$, the exact ground state of $\tilde{H}(\vec{Q})$. By projecting $\ket{\Psi_{ph}^{(1)}(\vec{Q})}$ onto $\ket{\phi_{0}(\vec{Q})}$, we can generally decompose it as:
\begin{equation}
    \ket{\Psi_{ph}^{(1)}(\vec{Q})} = a_{\vec{Q}}\ket{\phi_{0}(\vec{Q})}+b_{\vec{Q}}\ket{\chi(\vec{Q})},
\end{equation}
where the unknown wavefunction $\ket{\chi(\vec{Q})}$ is orthogonal to $\ket{\phi_{0}(\vec{Q})}$, i.e. $\braket{\phi_{0}(\vec{Q})|\chi(\vec{Q})}=0$, and $|a_{\vec{Q}}|^{2}+|b_{\vec{Q}}|^{2}=1$. Now, let us examine the ground state energy:
\begin{align}\label{eq:lowest_expectation_vs_q}
\begin{split}
    E_{\text{ph},1}(\vec{Q}) &= 
    \bra{\Psi_{ph}^{(1)}(\vec{Q})}\tilde{H}(\vec{Q})\ket{\Psi_{ph}^{(1)}(\vec{Q})} \\
    &= 
    \left[a_{\vec{Q}}^{*}\bra{\phi_{0}(\vec{Q})}+b_{\vec{Q}}^{*}\bra{\chi(\vec{Q})}\right]
    \left(\tilde{H}_{0}(\vec{Q})+\tilde{H}_{1}(\vec{Q})\right)
    \left[a_{\vec{Q}}\ket{\phi_{0}(\vec{Q})}+b_{\vec{Q}}\ket{\chi(\vec{Q})}\right] \\
    &=
    |a_{\vec{Q}}|^{2}\bra{\phi_{0}(\vec{Q})}
    \tilde{H}_{1}(\vec{Q})\ket{\phi_{0}(\vec{Q})}
    +|b_{\vec{Q}}|^{2}\bra{\chi(\vec{Q})}\left(\tilde{H}_{0}(\vec{Q})+\tilde{H}_{1}(\vec{Q})\right)\ket{\chi(\vec{Q})} +2\Re\lbrace b_{\vec{Q}}^{*}a_{\vec{Q}}\bra{\chi(\vec{Q})}\tilde{H}_{1}(\vec{Q})\ket{\phi_{0}(\vec{Q})}\rbrace.
\end{split}
\end{align}
Taking the limit $|\vec{Q}|/k_0\rightarrow 0$, and using \eqref{eq:exchange_of_lune}, we know that all matrix elements of $\tilde{H}_{1}(\vec{Q})$ vanish. The only term that does not necessarily vanish is:
\begin{equation}\label{eq:limit_e_vs_q}
    \lim_{|\vec{Q}|/k_0\rightarrow 0}E_{\text{ph},1}(\vec{Q}) = \lim_{|\vec{Q}|/k_0\rightarrow 0}|b_{\vec{Q}}|^{2}\bra{\chi(\vec{Q})}\tilde{H}_{0}(\vec{Q})\ket{\chi(\vec{Q})}.
\end{equation}
Using the same method as in Eqs.~\eqref{eq:lower_bound_derivation_q0}
--
\eqref{eq:lower_bound_wavefunction_q0}, we can write the lower bound for the expectation value in Eq.~\eqref{eq:limit_e_vs_q} as:
\begin{equation}
    \bra{\chi(\vec{Q})}\tilde{H}_{0}(\vec{Q})\ket{\chi(\vec{Q})} \ge |D_{\vec{Q}}|\min_{\substack{\vec{k},\vec{k}'\in D_{\vec{Q}}\\\vec{k}\neq\vec{k}'}}\lbrace-[\tilde{H}(\vec{Q}=0)]_{\vec{k},\vec{k}'}\rbrace.
\end{equation}
Inserting this bound back into Eq.~\eqref{eq:limit_e_vs_q}, we find:
\begin{equation}
    \lim_{|\vec{Q}|/k_0\rightarrow 0}E_{\text{ph},1}(\vec{Q}) \ge \frac{1}{4}e^{2}k_0\lim_{|\vec{Q}|/k_0\rightarrow 0}|b_{\vec{Q}}|^{2}.
\end{equation}
Combined with the upper bound on the lowest excitation energy found in Eq.~\eqref{eq:upper_bound_on_dispersion_vs_Q}, we find that:
\begin{equation}
    \lim_{|\vec{Q}|/k_0\rightarrow 0}\frac{E_{\text{ph},1}(\vec{Q})}{|\vec{Q}|/k_0}\leq \frac{4}{\pi}E_{0}(k_0)
    \implies \lim_{|\vec{Q}|/k_0\rightarrow 0}\frac{|b_{\vec{Q}}|^{2}}{|\vec{Q}|/k_0}\leq \frac{16}{\pi}\frac{E_{0}(k_0)}{e^{2}k_0}
    \implies |b_{\vec{Q}}|\sim \sqrt{\frac{|\vec{Q}|}{k_0}}.
\end{equation}
We can now derive the lower bound. Continuing from Eq.~\eqref{eq:lowest_expectation_vs_q} and using the positive-definiteness of $\tilde{H}(\vec{Q})$:
\begin{align}
\begin{split}
    E_{\text{ph},1}(\vec{Q}) 
    &\ge
    |a_{\vec{Q}}|^{2}\bra{\phi_{0}(\vec{Q})}
    \tilde{H}_{1}(\vec{Q})\ket{\phi_{0}(\vec{Q})}
    +2\Re\lbrace b_{\vec{Q}}^{*}a_{\vec{Q}}\bra{\chi(\vec{Q})}\tilde{H}_{1}(\vec{Q})\ket{\phi_{0}(\vec{Q})}\rbrace \\
    &\ge \bra{\phi_{0}(\vec{Q})}
    \tilde{H}_{1}(\vec{Q})\ket{\phi_{0}(\vec{Q})}\left(1+\mathcal{O}\left(\frac{|\vec{Q}|}{k_0}\right)\right)
    +\mathcal{O}\left(\frac{|\vec{Q}|^{3/2}}{k_0^{3/2}}\log{\left(\frac{|\vec{Q}|}{k_0}\right)}\right) \\
    &= \frac{4 E_{0}(k_0)}{\pi k_0} |\vec{Q}| 
    +\mathcal{O}\left(\frac{|\vec{Q}|^{3/2}}{k_0^{3/2}}\log{\left(\frac{|\vec{Q}|}{k_0}\right)}\right),
\end{split}
\end{align}
where in the last equality we inserted the expectation value as calculated in Eq.~\eqref{eq:uniform_expectation_vs_q}.
Combining this lower bound with the upper bound in Eq.~\eqref{eq:upper_bound_on_dispersion_vs_Q}, we find that:
\begin{equation}
    \frac{4 E_{0}(k_0)}{\pi k_0} |\vec{Q}| 
    +\mathcal{O}\left(\frac{|\vec{Q}|^{3/2}}{k_0^{3/2}}\log{\left(\frac{|\vec{Q}|}{k_0}\right)}\right) \leq E_{\text{ph},1}(\vec{Q}) \leq \frac{4 E_{0}(k_0)}{\pi k_0} |\vec{Q}| 
    +\mathcal{O}\left(\frac{|\vec{Q}|^{2}}{k_0^{2}}\log{\left(\frac{|\vec{Q}|}{k_0}\right)}\right),
\end{equation}
which implies that:
\begin{equation}
    E_{\text{ph},1}(\vec{Q}) \approx \frac{4 E_{0}(k_0)}{\pi k_0} |\vec{Q}|.
\end{equation}
For $k_0d\rightarrow \infty$, the velocity of these excitations is given by $v=\left(\frac{2}{\pi}\right)^{2}e^{2}=\left(\frac{2}{\pi}\right)^{2}\alpha c$, where $\alpha\approx\frac{1}{137}$ is the fine-structure constant and $c$ is the speed of light. This linear slope is added to Fig.~\ref{fig:ph_spectrum} by evaluating $E_{0}(k_0)$ for a finite system, and is found to be in agreement with the numerically computed excitation spectrum.

The isotropic result above is derived for the idealized dispersion in Eq.~\eqref{eq:well_dispersion}. One may naturally wonder what would happen upon the introduction of single-particle dispersion to the disk. Would we still have a linearly dispersive magnon? If so, what would be the velocity? Would it remain isotropic?

Let us assume our Hamiltonian is now of the form:
\begin{equation}
    \mathcal{H}_{\text{int}}\rightarrow \mathcal{H}_{\text{int}}+\sum_{\sigma=\uparrow,\downarrow}\sum_{|\vec{k}|\leq 
 k_0}E_{\text{disp}}(\vec{k})c_{\sigma,\vec{k}}^{\dagger}c_{\sigma,\vec{k}},
\end{equation}
where $E_{\text{disp}}(\vec{k})$ is a dispersion that is differentiable on the disk with a bounded gradient. Note that $E_{0}(\vec{k})$ in Eq.~\eqref{eq:exact_h_dispersion} for the case on an unscreened Coulomb interaction ($k_0d\rightarrow\infty$) does not possess this property, as its derivative diverges as $|\vec{k}|\rightarrow k_0$. We will address this dispersion separately. The Hamiltonian $\tilde{H}(\vec{Q})$ defined in Eq.~\eqref{eq:collective_matrix_shifted} acquires a new correction,
\begin{align}
\begin{split}
    \tilde{H}_{2}(\vec{Q})&=\sum_{\vec{k}\in D_{\vec{Q}}}\left(E_{\text{disp}}(\vec{k}+\vec{Q})-E_{\text{disp}}(\vec{k})\right)\ket{\vec{k}}\bra{\vec{k}} \\
    &\approx \sum_{\vec{k}\in D_{\vec{Q}}}\vec{v}_{\vec{k}}\cdot\vec{Q}\ket{\vec{k}}\bra{\vec{k}} +\mathcal{O}\left(\frac{|\vec{Q}|^{2}}{k_0^{2}}\right),
\end{split}
\end{align}
such that $\tilde{H}(\vec{Q})=\tilde{H}_{0}(\vec{Q})+\tilde{H}_{1}(\vec{Q})+\tilde{H}_{2}(\vec{Q})$, and we defined $\vec{v}_{\vec{k}}=\frac{\partial E_{\text{disp}}(\vec{k})}{\partial {\vec{k}}}$.
Following the previous arguments, the upper bound on the energy is given by:
\begin{align}
\begin{split}
    E_{\text{ph},1}(\vec{Q})&\leq \bra{\phi_{0}(\vec{Q})}\tilde{H}(\vec{Q})\ket{\phi_{0}(\vec{Q})} \\
    &= \bra{\phi_{0}(\vec{Q})}(\tilde{H}_{1}(\vec{Q})+\tilde{H}_{2}(\vec{Q}))\ket{\phi_{0}(\vec{Q})}
    \\
    &\approx \frac{4 E_{0}(k_0)}{\pi k_0} |\vec{Q}| 
    + \frac{1}{|D_{0}|}\sum_{\vec{k}\in D_{0}}\vec{v}_{\vec{k}}\cdot\vec{Q}
    +\mathcal{O}\left(\frac{|\vec{Q}|^{2}}{k_0^{2}}\log{\left(\frac{|\vec{Q}|}{k_0}\right)}\right).
\end{split}
\end{align}
The lower bound is derived similarly,
\begin{align}
\begin{split}
    E_{\text{ph},1}(\vec{Q})&\geq  \frac{4 E_{0}(k_0)}{\pi k_0} |\vec{Q}| 
    + \frac{1}{|D_{0}|}\sum_{\vec{k}\in D_{0}}\vec{v}_{\vec{k}}\cdot\vec{Q}
    +\mathcal{O}\left(\frac{|\vec{Q}|^{3/2}}{k_0^{3/2}}\log{\left(\frac{|\vec{Q}|}{k_0}\right)}\right),
\end{split}
\end{align}
such that the particle-hole dispersion is given by:
\begin{equation}
    E_{\text{ph},1}(\vec{Q}) \approx \frac{4 E_{0}(k_0)}{\pi k_0} |\vec{Q}| 
    + \frac{1}{|D_{0}|}\sum_{\vec{k}\in D_{0}}\vec{v}_{\vec{k}}\cdot\vec{Q}.
\end{equation}
We find that the smooth single-particle dispersion adds its average velocity on the disk to the isotropic velocity given by the strict confinement of the Coulomb interaction. Therefore, our result in Eq.~\eqref{eq:linear_ph_disp} holds for any smooth dispersion whose average velocity on the disk is zero.

Finally, consider adding the single-particle dispersion $E_{0}(\vec{k})$ in Eq.~\eqref{eq:exact_h_dispersion}, such that:
\begin{align}
\begin{split}
    \tilde{H}_{2}(\vec{Q})&=\sum_{\vec{k}\in D_{\vec{Q}}}\left(E_{0}(\vec{k}+\vec{Q})-E_{0}(\vec{k})\right)\ket{\vec{k}}\bra{\vec{k}}.
\end{split}
\end{align}
The difficulty with blindly repeating the previous arguments occurs for the unscreened Coulomb interaction, where logarithmically singular terms of the form $\frac{|\vec{Q}|}{k_0}\log{\left(\frac{|\vec{Q}|}{k_0}\right)}$ appear in the small $|\vec{Q}|/k_0$ expansion of this Hamiltonian.
Using the rotational symmetry of $E_{0}(\vec{k})$ it is easy to show that, in this case, $\bra{\phi_{0}(\vec{Q})}\tilde{H}_{2}(\vec{Q})\ket{\phi_{0}(\vec{Q})}=0$. Consequently, the upper bound on $E_{\text{ph},1}(\vec{Q})$ is unchanged by this dispersion. The lower bound is also unchanged, as the logarithmically singular terms in $\tilde{H}_{2}(\vec{Q})$ play a similar role to such terms that are already present in $\tilde{H}_{1}(\vec{Q})$. We therefore conclude that that Eq.~\eqref{eq:linear_ph_disp} holds even when one adds the single-particle dispersion $E_{0}(\vec{k})$ to Eq.~\eqref{eq:well_dispersion}, i.e. for $\mathcal{H}_{\rho-\rho}$ as defined in Eq.~\eqref{eq:psd_decomposition}.

\subsection{The softened momentum confinement}
So far we have only considered the strict momentum confinement of Eq.~\eqref{eq:well_dispersion}. Now, we would like to consider a softened momentum confinement, $\mathcal{H}_{\text{K}}$, as defined in Eq.~\eqref{eq:hamiltonian_soft_well}, and calculate the dispersion of the particle-hole excitations above the same spin-polarized state $\ket{\Psi_{\text{SP}}}$. In Fig.~\ref{fig:ph_spectrum}(b) we plot the lowest lying excitation for various powers of $N_{D}$ (the strength of the soft confinement), and compare the result to the limit of $N_{D}\rightarrow\infty$ obtained above. We emphasize that the state $\ket{\Psi_{\text{SP}}}$ is not the ground state of the Hamiltonian $\mathcal{H}_{\text{K}}$, but for large enough $N_{D}$ we expect it to be a good approximation.

The softened momentum confinement means we allow particle excitations beyond the disk. Consequently, the Coulomb interaction is no longer restricted to the disk, thus we denote it by $\mathcal{V}_{\text{int}}$ to distinguish it from $\mathcal{H}_{\text{int}}$:
\begin{equation}
    \mathcal{V}_{\text{int}} 
    = \frac{1}{2A}\sum_{\sigma,\sigma'=\uparrow,\downarrow}\sum_{\vec{k},\vec{k}',\vec{q}}V_{\vec{q}}c_{\sigma',\vec{k}'-\vec{q}}^{\dagger}c_{\sigma,\vec{k}+\vec{q}}^{\dagger}c_{\sigma,\vec{k}}c_{\sigma',\vec{k}'}.
\end{equation}
The first step is to compute the matrix elements of $\mathcal{V}_{\text{int}}+\mathcal{H}_{\text{K}}$ between different wavefunctions $\ket{\Psi_{\text{ph}}^{\vec{k},\vec{Q}}}$. For the softened confinement, the momentum $\vec{k}$ is the entire disk for any $\vec{Q}$, and we find the following matrix for $\vec{k},\vec{k}'\in D_{0}$:
\begin{align}\label{eq:ph_hamiltonian_soft_well}
\begin{split}
    [H(\vec{Q})]_{\vec{k},\vec{k}'} 
    &= \bra{\Psi_{\text{ph}}^{\vec{k}',\vec{Q}}}(\mathcal{V}_{\text{int}}+\mathcal{H}_{\text{K}})
    \ket{\Psi_{\text{ph}}^{\vec{k},\vec{Q}}}
    \\
    &= \delta_{\vec{k},\vec{k}'}\left[E_{\text{SP}}+\frac{1}{A}\sum_{|\vec{p}|\leq k_0}V_{\vec{k}-\vec{p}}+U_{\text{K}}\left(\left(\frac{|\vec{k}+\vec{Q}|}{k_0}\right)^{N_{D}}-\left(\frac{|\vec{k}|}{k_0}\right)^{N_{D}}\right)\right]-\frac{1}{A}V_{\vec{k}-\vec{k}'}
\end{split}
\end{align}
We retain the notation of the Hamiltonian, the eigensystem and the basis (Eqs.~\eqref{eq:collective_matrix_shifted},\eqref{eq:ph_eigensystem} and Eq.~\eqref{eq:simplified_ph_basis_notation}, respectively), as the two problems can, for the most part, be considered aspects of the same one with a different choice of a parameter (finite vs infinite $N_{D}$). The only relevant difference is the number of finite eigenvalues, which is now unconstrained by $\vec{Q}$ (i.e., $n=1,\dots ,N$ where $N=|D_0|$):
\begin{equation}
    \tilde{H}(\vec{Q})\ket{\Psi_{ph}^{(n)}(\vec{Q})}=E_{\text{ph},n}(\vec{Q})\ket{\Psi_{ph}^{(n)}(\vec{Q})},\quad E_{\text{ph},1}(\vec{Q})\leq E_{\text{ph},2}(\vec{Q}) \leq \dots \leq E_{\text{ph},N}(\vec{Q}).
\end{equation}

We are interested in evaluating $E_{\text{ph},1}(\vec{Q})$ to leading order in $|\vec{Q}|/k_0$. To second order in perturbation theory, we find:
\begin{align}\label{eq:second_order_pert_eq}
\begin{split}
    E_{\text{ph},1}(\vec{Q}) &\approx E_{\text{ph},1}(0)
    +\bra{\Psi_{ph}^{(1)}(0)}\tilde{H}(\vec{Q})\ket{\Psi_{ph}^{(1)}(0)}
    +\sum_{n=2}^{N}\frac{\left|\bra{\Psi_{ph}^{(n)}(0)}\tilde{H}(\vec{Q})\ket{\Psi_{ph}^{(1)}(0)}\right|^{2}}{E_{\text{ph},1}(0)-E_{\text{ph},n}(0)},
\end{split}
\end{align}
where $\tilde{H}(\vec{Q})$ should be expanded up to second order in $|\vec{Q}|/k_0$.
The exact ground state of at $\vec{Q}=0$ is known exactly:
\begin{equation}
    \ket{\Psi_{ph}^{(1)}(0)} = \frac{1}{\sqrt{N}}\sum_{|\vec{k}|\leq k_0}\ket{\vec{k}},
\end{equation}
and we have $E_{\text{ph},1}(0)=\tilde{H}(0)\ket{\Psi_{ph}^{(1)}(0)}=0$. Therefore, the zeroth order in $|\vec{Q}|/k_0$ vanishes, as expected.
For the first and second order in $|\vec{Q}|/k_0$, we expand the Hamiltonian as:
\begin{align}
\begin{split}
    [\tilde{H}(\vec{Q})]_{\vec{k},\vec{k}'} 
    &\approx
    [\tilde{H}(0)]_{\vec{k},\vec{k}'}
    +\delta_{\vec{k},\vec{k}'}U_{\text{K}}\frac{N_{D}}{2}
    \left(\frac{|\vec{k}|}{k_0}\right)^{N_{D}-4}\frac{2|\vec{k}|^{2}\vec{k}\cdot\vec{Q}+|\vec{k}|^{2}|\vec{Q}|^{2}+(N_{D}-1)\left(\vec{k}\cdot\vec{Q}\right)^2}{k_0^{4}}+\mathcal{O}\left(\frac{|\vec{Q}|^3}{k_0^3}\right).
\end{split}
\end{align}
Applying this expansion to the exact ground state at $\vec{Q}=0$, we find:
\begin{align}
\begin{split}
    \tilde{H}(\vec{Q})\ket{\Psi_{ph}^{(1)}(0)} &\approx 
    \frac{1}{\sqrt{N}}U_{\text{K}}\frac{N_{D}}{2}
    \sum_{|\vec{k}|\leq k_0}\left(\frac{|\vec{k}|}{k_0}\right)^{N_{D}-4}
    \frac{2|\vec{k}|^{2}\vec{k}\cdot\vec{Q}+|\vec{k}|^{2}|\vec{Q}|^{2}+(N_{D}-1)\left(\vec{k}\cdot\vec{Q}\right)^2}{k_0^{4}}
    \ket{\vec{k}}.
\end{split}
\end{align}
Equipped with the result above, we can evaluate the perturbative corrections to the ground state energy.

The first order correction term, which has also $|\vec{Q}|^{2}/k_0^2$ terms, is given by:
\begin{align}
\begin{split}
    \bra{\Psi_{ph}^{(1)}(0)}\tilde{H}(\vec{Q})\ket{\Psi_{ph}^{(1)}(0)} &\approx 
    \frac{1}{N}U_{\text{K}}\frac{N_{D}}{2}
    \sum_{|\vec{k}|\leq k_0}\left(\frac{|\vec{k}|}{k_0}\right)^{N_{D}-4}
    \frac{2|\vec{k}|^{2}\vec{k}\cdot\vec{Q}+|\vec{k}|^{2}|\vec{Q}|^{2}+(N_{D}-1)\left(\vec{k}\cdot\vec{Q}\right)^2}{k_0^{4}} \\
    &\approx \frac{4\pi}{k_0^2}U_{\text{K}}\frac{N_{D}}{2}
    \int_{0}^{k_0}\frac{kdk}{2\pi}\int_{0}^{2\pi}\frac{d\theta}{2\pi}\left(\frac{k}{k_0}\right)^{N_{D}-4}
    \frac{k^{2}Q^{2}+(N_{D}-1)k^2Q^2\cos^{2}{\theta}}{k_0^{4}}\\
    &= \frac{1}{2}U_{\text{K}}\left(N_{D}+1\right)\frac{Q^2}{k_0^{2}},
\end{split}
\end{align}
where we used $\frac{N}{A}=\frac{k_0^{2}}{4\pi}$ as the density of electrons in the system.

The second order correction is strictly negative. We will now bound its magnitude.
For $\vec{Q}=0$, the finite and infinite $N_{D}$ cases coincide, as we consider the same ground state. Therefore, the energy spectrum is bounded between the same two values as in Eq.~\eqref{eq:bound_on_ph_energy}:
\begin{align}
    \frac{1}{2e^{2}k_0}
    \leq\frac{1}{|E_{\text{ph},1}(0)-E_{\text{ph},n}(0)|}
    \leq \frac{4}{e^{2}k_0} 
\end{align}
Therefore, up to a constant $\frac{1}{2}\leq C \leq 4$, we find the second order correction to be:
\begin{align}
\begin{split}
    \sum_{n=2}^{N}\frac{\left|\bra{\Psi_{ph}^{(n)}(0)}\tilde{H}(\vec{Q})\ket{\Psi_{ph}^{(1)}(0)}\right|^{2}}{|E_{\text{ph},1}(0)-E_{\text{ph},n}(0)|} &= \frac{C}{e^{2}k_0} \sum_{n=2}^{N}\left|\bra{\Psi_{ph}^{(n)}(0)}\tilde{H}(\vec{Q})\ket{\Psi_{ph}^{(1)}(0)}\right|^{2}
    \\
    &= \frac{C}{e^{2}k_0} \left(\left|\left|\tilde{H}(\vec{Q})\ket{\Psi_{ph}^{(1)}(0)}\right|\right|^{2}-\left|\bra{\Psi_{ph}^{(1)}(0)}\tilde{H}(\vec{Q})\ket{\Psi_{ph}^{(1)}(0)}\right|^{2}\right)
\end{split}
\end{align}
Including terms up to second order in $|\vec{Q}|/k_0$, only the first term has a $Q^{2}$ component:
\begin{align}
\begin{split}
    \sum_{n=2}^{N}\frac{\left|\bra{\Psi_{ph}^{(n)}(0)}\tilde{H}(\vec{Q})\ket{\Psi_{ph}^{(1)}(0)}\right|^{2}}{|E_{\text{ph},1}(0)-E_{\text{ph},n}(0)|} &\approx \frac{C}{N}\frac{U_{\text{K}}^{2}}{e^{2}k_0}N_{D}^{2}\sum_{|\vec{k}|\leq k_0}
    \left(\frac{|\vec{k}|}{k_0}\right)^{2N_{D}-4}
    \frac{(\vec{k}\cdot\vec{Q})^{2}}{k_0^{4}} \\
    &\approx
    \frac{C}{2}\frac{U_{\text{K}}^{2}}{e^{2}k_0}N_{D}\frac{Q^2}{k_0^{2}}
\end{split}
\end{align}
where $\frac{1}{2}\leq C \leq 4$.

Taking $N_{D}\gg1$, we find that for small enough $Q/k_0$:
\begin{equation}
    E_{\text{ph},1}(\vec{Q})\approx \frac{1}{2}U_{\text{K}}\left(1-C\frac{U_{\text{K}}}{e^{2}k_0}\right)N_{D}\frac{Q^{2}}{k_0^{2}}
\end{equation}
Therefore, if $U_{\text{K}}$ is small compared to the next particle-hole excitation energy, i.e. $U_{\text{K}}\ll e^{2}k_0$, the second term in the parenthesis becomes unimportant, and we obtain the positive and divergent spin stiffness in Eq.~\eqref{eq:soft_ph_disp}. The divergence can be seen numerically in Fig.~\ref{fig:ph_spectrum} where $E_{\text{ph},1}(\vec{Q})$ is computed numerically for several values of $N_{D}$. We remind the reader that while the particle-hole excitation here is calculated with respect to a state that is not the ground state of the system, we expect that the statement above will be qualitatively correct for large enough $N_{D}$.

\newpage
\section{Numerical Analysis}\label{sec:numerics}

\subsection{Discretization of momentum space}
We begin by describing our finite-size model. Consider a finite 2D system of area $A$ with periodic boundary conditions. In an effort to utilize as many symmetries as possible, we choose the simulation box to be invariant under $\frac{\pi}{3}$ rotations. In real space, this box is a rhombus with an angle of $\frac{\pi}{3}$ and a side of length $L=\sqrt{\frac{2}{\sqrt{3}}A}$. Correspondingly, the momentum space is discretized as a triangular lattice with periodicity $\Delta k = \frac{4\pi}{\sqrt{3}L}$. We are interested in considering only momenta up to a cutoff $k_0$. By adjusting the area $A$, we set $k_0=R\cdot \Delta k$ for some value of the dimensionless parameter $R$ which determines the number of grid points up to the cutoff. The system area is therefore set to be $A=\frac{2}{\sqrt{3}}\frac{(2\pi)^{2}}{k_0^{2}}R^2$. The smallest values of $R$ for which $k_0$ sits on the reciprocal lattice are $R=1$, $R=\sqrt{3}$ and $R=2$. For $1\leq R\leq 6$ where $R$ is an integer, the number of grid points is $N=3R^2+3R+1$, such that at $R=1$ and $R=2$ we obtain $N=7$ and $N=19$ grid points, respectively. The intermediate value of $R=\sqrt{3}$ gives $N=13$. In the thermodynamic limit (taking $A\rightarrow\infty$, which implies $R\rightarrow\infty$), we find $N\approx \frac{2\pi}{\sqrt{3}} R^{2}$. For $R > 2$ the number of grid points is too large for exact diagonalization, as the Hilbert space dimension is exponentially large. The particle-hole excitation calculation has a Hilbert space dimension that is linear in $N$, and there we use $R=6$. 
At filling factor $\nu=1$ we place one electron at each grid point, and the electron density $n=\frac{N}{A}$ is given by:
\begin{equation}\label{eq:finite_system_density_factor}
    \frac{N}{A}=
    \begin{cases}
        \frac{\sqrt{3}}{2}\frac{3R^2+3R+1}{R^{2}}\frac{k_0^{2}}{(2\pi)^{2}}, & R \in \lbrace1,2,3,4,5,6\rbrace \\
        \frac{\sqrt{3}}{2}\frac{13}{3}\frac{k_0^{2}}{(2\pi)^{2}}, & R=\sqrt{3} \\
        \pi\frac{k_0^{2}}{(2\pi)^{2}}, & R\rightarrow\infty.
    \end{cases}
\end{equation}
We use the density factor obtained for small $R$ when presenting numerical results in Fig.~\ref{fig:egs_vs_meff}, and the one for infinite $R$ when discussing analytical results.

In the discretization above, the periodic boundary conditions implied the number of grid points is always odd. In order to simulate an even number of grid points, we introduce twisted boundary conditions on the real space rhombus, identifying pairs of opposite edges with a phase factor of $e^{i2\pi/3}$. This choice of phase factor preserves $C_{3}$ rotation symmetry in momentum space, with the rotation axis being the center of an equilateral triangle of reciprocal lattice points. The smallest values for $R$ in this case are $R=\sqrt{\frac{1}{3}}$, $R=\sqrt{\frac{4}{3}}$, $R=\sqrt{\frac{7}{3}}$, and $R=\sqrt{\frac{13}{3}}$, which correspond to $N=3$, $N=6$, $N=12$ and $N=18$, respectively. Further increasing $R$ results in a Hilbert space dimension too large for exact diagonalization. The electron density at filling factor $\nu=1$ is calculated in a similar manner to Eq.~\eqref{eq:finite_system_density_factor}, and we obtain:
\begin{equation}
    \frac{N}{A}=
    \begin{cases}
        \frac{\sqrt{3}}{2}\frac{3\cdot 3}{1}\frac{k_0^{2}}{(2\pi)^{2}}, & R=\sqrt{\frac{1}{3}}\quad(N=3)\\
        \frac{\sqrt{3}}{2}\frac{3\cdot 6}{4}\frac{k_0^{2}}{(2\pi)^{2}}, & R=\sqrt{\frac{4}{3}}\quad(N=6)\\
        \frac{\sqrt{3}}{2}\frac{3\cdot 12}{7}\frac{k_0^{2}}{(2\pi)^{2}}, & R=\sqrt{\frac{7}{3}}\quad(N=12)\\
        \frac{\sqrt{3}}{2}\frac{3\cdot 18}{13}\frac{k_0^{2}}{(2\pi)^{2}}, & R=\sqrt{\frac{13}{3}}\quad(N=18).
    \end{cases}
\end{equation}
In this work we present and discuss simulations starting with $N=6$ and up to $N=19$ grid points.

\subsection{Exact diagonalization}
The numerical calculations were performed on the WEXAC cluster at the Weizmann Institute of Science. The exact diagonalization is implemented using the matrix-free implicitly restarted Lanczos method, converging $7$ lowest eigenvalues to an accuracy of $10^{-5}$. The algorithm is implemented using the \texttt{ARPACK}~\cite{Lehoucq1998-ew} library as maintained by \texttt{opencollab ARPACK-NG}~\cite{Opencollab_undated-vd}. 

\subsubsection{Procedure to obtain the ground state's energy and spin polarization}
Next, we describe the procedure used to exactly diagonalize the Hamiltonians in Eq.~\eqref{eq:single_orbital_hamiltonian} and Eq.~\eqref{eq:massive_hamiltonian} for the finite system of $N$ grid points defined above. We are interested in finding the ground state energy at a filling factor of $\nu=1$, determining its total spin, and repeating the calculation for different effective masses. The first step is to separate the Hilbert space into small sectors that are invariant subspaces of the Hamiltonian. This reduction in size will render the exact diagonalization tractable. The second step will be to extract the information about the spin polarization.

The entire Hilbert space $\mathcal{H}$ associated with the Hamiltonians of interest consists of two orbitals at each grid point, one for each spin projection, such that $\dim \mathcal{H} = 2^{2N}$. Focusing on a fixed number of particles equal to the number of grid points $N$, i.e. for filling factor $\nu=1$, the relevant Hilbert space is reduced to $\dim \mathcal{H} = {{2N}\choose{N}}$. We can further decrease our Hilbert space dimension by considering sectors of total momentum, $\vec{k}_{\text{tot}}=\sum_{\sigma}\sum_{|\vec{k}|\leq k_0}\vec{k}c_{\sigma,\vec{k}}^{\dagger}c_{\sigma,\vec{k}}$ , and of total spin projection, $S_{z}=\frac{1}{2}\sum_{|\vec{k}|\leq k_0}(c_{\uparrow,\vec{k}}^{\dagger}c_{\uparrow,\vec{k}}-c_{\downarrow,\vec{k}}^{\dagger}c_{\downarrow,\vec{k}})$. The largest sector is the $\left(\vec{k}_{\text{tot}},S_{z}\right)=\left(0,\frac{1}{2}\right)$ sector for $N=19$ which has $\dim\mathcal{H}\approx 9\cdot 10^{7}$. While still very large, this sector is small enough to allow for its exact diagonalization.

There are more symmetries we wish to exploit - the point group symmetries of the hexagonal lattice (rotation and mirror), and the $SU(2)$ symmetry associated with rotations of the total spin. These symmetries have important yet very different consequences on our numerical calculations. 
The point group symmetries imply that there is complete redundancy in information from different sectors of $\vec{k}_{\text{tot}}$ that are related by symmetry, thereby the number of distinct $\vec{k}_{\text{tot}}$ sectors we need to consider is significantly reduced. In a similar manner, the spin rotation symmetry allows us to consider only spin sectors with $S_{z}\ge 0$, as the negative values of $S_{z}$ contain the same information as their positive counterparts. 

The $SU(2)$ symmetry implies that the total spin $S$ and the total spin projection $S_{z}$ are good quantum numbers. With that said, constructing the sectors of fixed $S,S_z$ can be quite difficult. However, constructing a sector of given $S_{z}$ is computationally easy.
Given that our goal is to find the spin polarization of a state, this might seem insufficient, but as we now show,
the spin rotational degree of freedom allows us to explore the total spin of the eigenstates. To see that, consider a state with total spin $S$. Due to $SU(2)$ symmetry, this state will be $2S+1$ degenerate and will appear in sectors with $S_{z}=-S,\dots,S$. Since total spin is a conserved number, any eigenstate with an odd (even) number of particles necessarily is degenerate with a representative in the $S_{z}=\frac{1}{2}$ ($S_{z}=0$) sector. Therefore, the $S_{z}=\frac{1}{2}$ ($S_{z}=0$) sector contains the entire eigenvalue spectrum for an odd (even) number of particles. Sweeping different $S_{z}$ sectors allows us to identify the total spin $S$ of every state as the maximal $S_{z}$ sector in which it appears. 

We summarize the argument above with the following recipe to find the ground state and its spin polarization for $N$ particles: 
\begin{enumerate}
    \item List all sectors of $S_{z}=\frac{(N\mod 2)}{2}$ and $\vec{k}_{\text{tot}}$ that are unrelated by the point group symmetries.
    \item Diagonalize each sector to find its lowest eigenvalue.
    \item Identify the momentum sector hosting the lowest eigenvalue across all the sectors. This momentum is the ground state momentum (up to point group symmetries), and the eigenvalue is the ground state energy.
    \item Repeat the exact diagonalization for $S_{z}=S_{z}+1$ only in the $\vec{k}_{\text{tot}}$ sector that hosts the ground state.
    \item The spin polarization $S$ of the ground state is the maximal value of $S_{z}$ that displays the ground state energy found at $S_{z}=\frac{(N\mod 2)}{2}$.
\end{enumerate}

\subsubsection{Simulation results}
The results of the first and second steps of the procedure described above are shown in Fig.~\ref{fig:egs_vs_ktot}, where the lowest energy eigenvalue of the Hamiltonian in Eq.~\eqref{eq:massive_hamiltonian} as obtained by exact diagonalization is plotted as a function of the momentum sector $\vec{k}_{\text{tot}}$ (and fixed $S_{z}=\frac{(N\mod 2)}{2}$). Since the Hamiltonian in Eq.~\eqref{eq:massive_hamiltonian} has an effective mass as a parameter, we repeat this simulation for a range of inverse effective masses $m_{\text{eff}}^{-1}$, increased in  steps of size $m_0^{-1}=\frac{V_{k_0}N}{2A k_0^{2}}$. For the special value of $m_{\text{eff}}^{-1}=0$ where the Hamiltonian in  Eq.~\eqref{eq:massive_hamiltonian} is reduced to Eq.~\eqref{eq:single_orbital_hamiltonian} we find the relation between energy and momentum is approximately linear. This linearity is due to the single velocity scale ($E/k_0\sim e^2$) in this Hamiltonian.

Proceeding with steps (3)-(5), we identify the momentum sector hosting the lowest energy (for each $m_{\text{eff}}^{-1}$) and repeat the diagonalizing in this sector for different values of $S_{z}$ to identify its total spin. The results for the ground state momentum, energy and total spin are shown in Fig~\ref{fig:egs_vs_meff}.
We find that the ground state for $-3m_0^{-1} \lesssim m_{\text{eff}}^{-1}\lesssim 2m_0^{-1}$, is fully spin polarized (which implies it has $\vec{k}_{\text{tot}}=0$) for all system sizes, with the precise polarized phase boundaries slightly varying with  system size (see Fig.~\ref{fig:spin_vs_m} in the main text). 
Beyond the spin polarized regime, i.e. for $m_{\text{eff}}^{-1}\lesssim-3m_0^{-1}$ or $2m_0^{-1}\lesssim m_{\text{eff}}^{-1}$, we find the ground state has a total spin and total momentum which varies with system size. 

At first sight, in the range of $m_{\text{eff}}^{-1}\lesssim-3m_0^{-1}$ or $2m_0^{-1}\lesssim m_{\text{eff}}^{-1}$, the ground state seems (for some system sizes) to be partially spin-polarized and for some cases to also have a non-zero momentum. However, we now argue that this is a finite-size artifact. To make the discussion concrete, we will consider the case of $N=19$, but the following arguments hold with trivial modifications to every $N$. In the limit of $m_{\text{eff}}^{-1}\rightarrow \pm\infty$, we expect a completely depolarized state, i.e. $S_{z}=\frac{1}{2}$. This is because the Coulomb interaction is completely negligible compared to the kinetic dispersion. In this limit, Hund's rule dictates that the favorable configuration among states with the same kinetic energy is the one that maximizes the total spin. Consider placing $N=19$ electrons on our momentum grid one at a time while minimizing kinetic energy. For $m_{\text{eff}}^{-1}\rightarrow \infty$ ($m_{\text{eff}}^{-1}\rightarrow -\infty$) we find that the last $5$ electrons ($7$ electrons) have to be placed on $6$ spin-degenerate sites with momenta $k=\frac{\sqrt{3}}{2}k_0$ (see Fig.~\ref{fig:hund}). This leads to a ${12}\choose{5}$ degenerate ground state for either sign of $m_{\text{eff}}^{-1}$. In this subspace of ground states, the states with maximal total spin have $S=\frac{5}{2}$, and a $12$-fold degeneracy due to $C_{6}$ rotations and $S_{z}=\pm\frac{5}{2}$. All the maximal spin configurations occur with $|\vec{k}_{\text{tot}}|=\frac{\sqrt{3}}{2}k_0$. By fixing $\vec{k}_{\text{tot}}$ and $S_{z}>0$, it follows from Hund's rule that the ground state has this total spin and momentum. A similar argument for the $N=7$ system leads to $S=\frac{5}{2}$ with $|\vec{k}_{\text{tot}}|=k_0$, in perfect agreement with our numerical results. In Fig.~\ref{fig:hund} we also describe this schematically for $N=18$. We expect that increasing the system size will retain this residual spin polarization which gradually becomes insignificant: $S/N\rightarrow0$. In the thermodynamic limit, this artifact would not occur. This is hinted in Fig.~\ref{fig:spin_vs_m} by considering the total spin per particle for different system sizes. We note that for $N=19$ particles, if we were to focus our attention on the $\vec{k}_{\text{tot}}=0$ sector where the thermodynamic limit's ground state is expected, we find (numerically) its lowest energy state has $S=\frac{3}{2}$, as expected from the exact same Hund's rule argument above. We conclude that the phase outside $-3m_0^{-1} \lesssim m_{\text{eff}}^{-1}\lesssim 2m_0^{-1}$ is consistent with a spin unpolarized phase up to finite-size effects.

\newpage
\begin{figure}
    \centering
    \includegraphics[width=\textwidth]{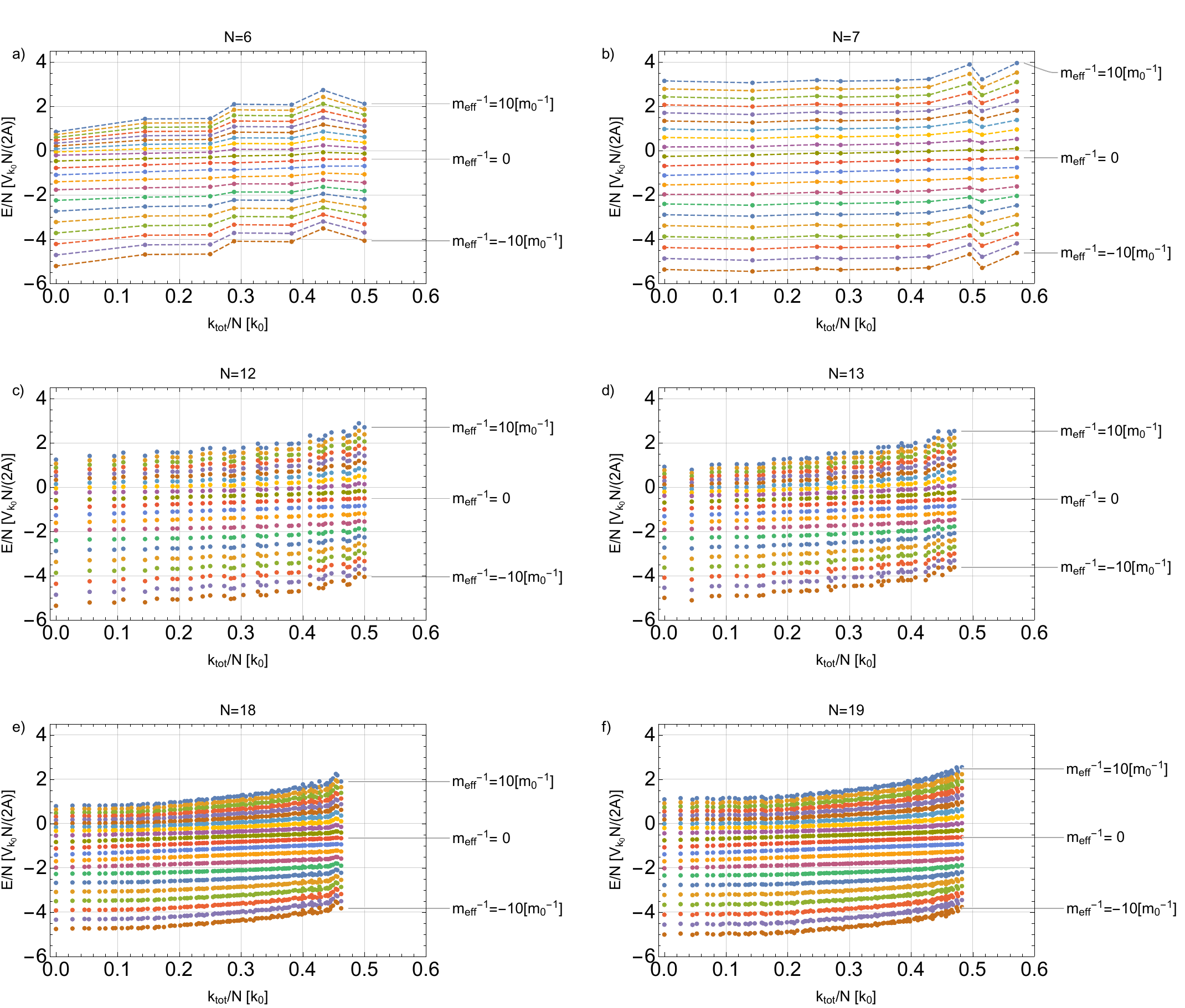}
    \caption{The lowest energy per particle of Eq.~\eqref{eq:massive_hamiltonian} at filling factor of $\nu=1$, as obtained by exact diagonalization, as a function of total momentum sector $|\vec{k}_{\text{tot}}|$, and for a range of inverse effective masses $m_{\text{eff}}^{-1}$. Figures (a)-(f) correspond to $N=6,7,12,13,18,19$} grid points, respectively.
    \label{fig:egs_vs_ktot}
\end{figure}

\newpage
\begin{figure}
    \centering
    \includegraphics[width=\textwidth]{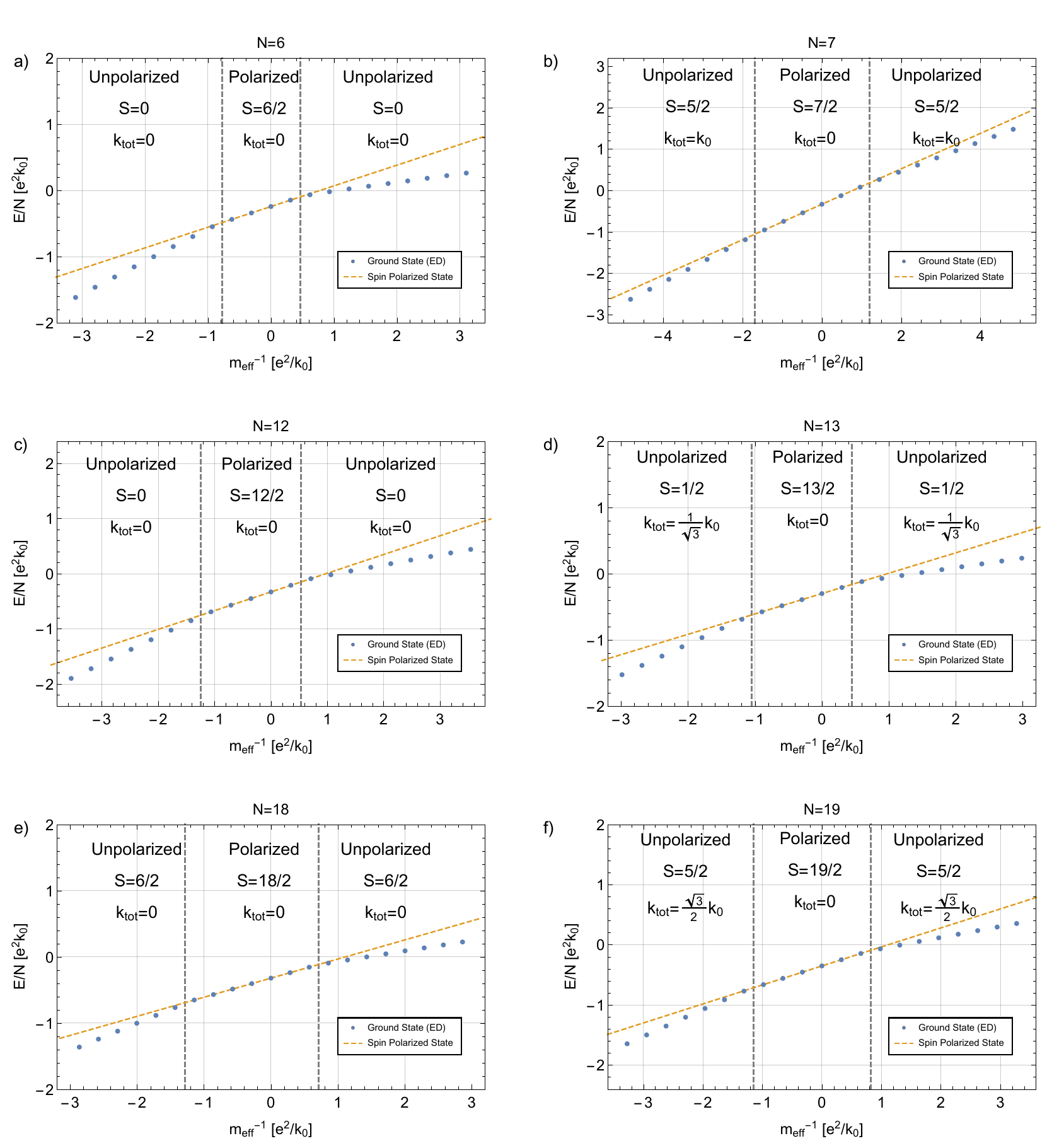}
    \caption{The ground state energy per particle of Eq.~\eqref{eq:massive_hamiltonian} at filling factor of $\nu=1$, as obtained by exact diagonalization, as a function of inverse effective mass. Figures (a)-(f) correspond to $N=6,7,12,13,18,19$ grid points, respectively. The dashed yellow line corresponds to the energy of the spin-polarized state, which exactly matches the ground state for data points between the vertical dashed lines. The overlayed text specifies the total spin and momentum of the obtained ground state.}
    \label{fig:egs_vs_meff}
\end{figure}

\newpage
\begin{figure}
    \centering
    \includegraphics[width=\textwidth]{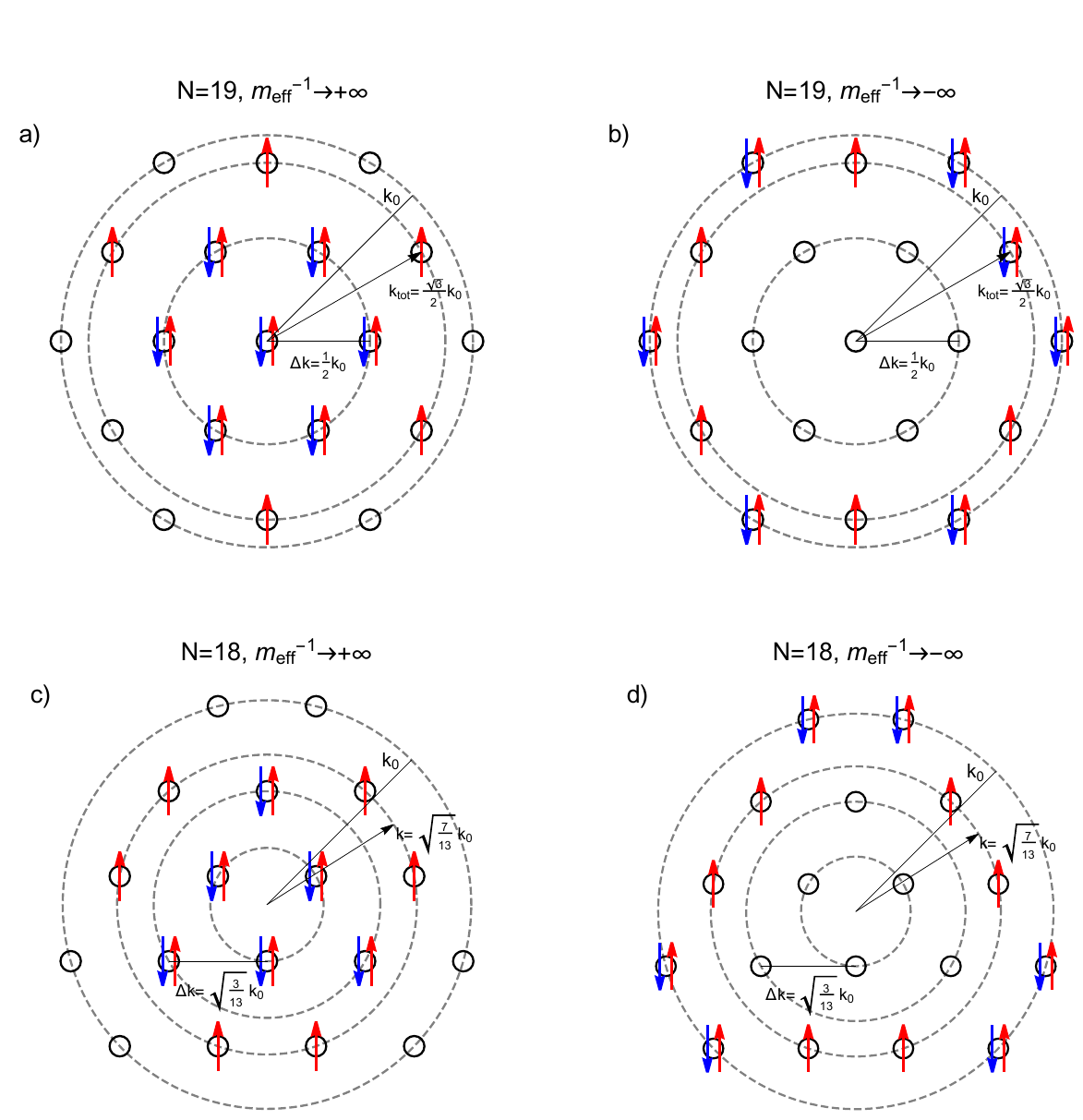}
    \caption{The ground state spin configuration of the Hamiltonian in Eq.~\eqref{eq:massive_hamiltonian} at filling factor $\nu=1$ for $N=18$ and $N=19$ in the limit of $m_{\text{eff}}^{-1}\rightarrow \pm\infty$, as dictated by Hund's rule. The small solid circles denote the momentum grid, while the dashed concentric circles are shells that connect sites with degenerate kinetic energy. For $N=19$, the partially filled shell of momentum $k=\frac{\sqrt{3}}{2}k_0$ is occupied in the configuration that maximizes the total spin. This configuration is found to have a total spin of $S=\frac{5}{2}$ and a total momentum (indicated by the black arrow) of $|\vec{k}_{\text{tot}}|=\frac{\sqrt{3}}{2}k_0$. For $N=18$, the partially filled shell of momentum $k=\sqrt{\frac{7}{13}}k_0$ is occupied in the configuration that maximizes the total spin. In this case, the total spin is $S=6/2$ and the total momentum is $|\vec{k}_{\text{tot}}|=0$.}
    \label{fig:hund}
\end{figure}

\end{widetext}
\end{document}